\documentclass[amsmath,amssymb,twocolumn,superscriptaddress]{revtex4}

\usepackage{graphicx}
\usepackage{dcolumn}

\def\be{\begin{equation}}
\def\ee{\end{equation}}
\def\beq{\begin{eqnarray}}
\def\eeq{\end{eqnarray}}

\usepackage{bm}
\usepackage{graphicx}
\usepackage{xspace}
\usepackage{dcolumn}
\usepackage{amsmath}
\usepackage[latin1]{inputenc}
\usepackage{graphicx, psfrag}
\usepackage{amssymb}
\usepackage[colorlinks=true, citecolor=blue, urlcolor = blue, linkcolor= red, bookmarks=true]{hyperref}
\usepackage{float}
\usepackage{amsmath}
\usepackage{amsfonts}
\usepackage{dcolumn}
\usepackage{hyperref}
\usepackage{subfigure}
\usepackage{pgfplots}
\usepackage{epstopdf}
\usepackage{booktabs}
\usepackage{multirow}
\usepackage{ctable}
\usepackage{tabularray}

\begin{document}



\title{ Study on physical properties and characteristics of an anisotropic compact star model using Karmarkar Condition in $\mathcal{F}(\mathcal{Q})$ gravity} 

\author{Sat Paul}
\email[Email:]{satpaul232039cuh@gmail.com}
\affiliation{Department of Mathematics, Central University of Haryana, Jant-Pali, Mahendergarh 123029, Haryana, India}

\author{Jitendra Kumar}
\email[Email:]{jitendark@gmail.com}
\affiliation{Department of Mathematics, Central University of Haryana, Jant-Pali, Mahendergarh 123029, Haryana, India}  

\author{S. K. Maurya}
\email[Email:]{sunil@unizwa.edu.om}
\affiliation{Department of Mathematical and Physical Sciences,
College of Arts and Sciences, University of Nizwa, P.O. Box 33, Nizwa 616, Sultanate of Oman}
\affiliation{Research Center of Astrophysics and Cosmology, Khazar University, Baku, AZ1096, 41 Mehseti Street, Azerbaijan}

\date{\today}

\begin{abstract} The main aim of this study is to examine the behaviour of physical parameters of an anisotropic compact star model demonstrating spherical symmetry in $\mathcal{F}(\mathcal{Q})$ modified gravity. To evaluate the behaviour and the stability of an anisotropic compact star model, we utilise the measured mass and radius of an anisotropic compact star model. This study obtained an anisotropic compact star model by solving Einstein's field equations. The field equations have been simplified by an appropriate selection of the metric element $e^{\sigma}$ and the Karmarkar condition. By solving field equation to develop a differential equation that establishes a relationship between two essential components of spacetime, namely $e^{\sigma}$ and $e^{\Upsilon}$. A physical analysis of this model reveals that the resulting stellar structure for anisotropic matter distribution is a physically plausible representation of a compact star with an energy density of order $10^{14}g/cm^{3}$. Using the Tolman-Oppenheimer-Volkoff equation, causality condition and Harrison-Zeldovich-Novikov Condition, we investigate the hydrostatic equilibrium and stability of the compact star Cen X-3. We further determined the mass-radius relation of this compact star for different values of $\delta_{1}$.
 \end{abstract}
\maketitle
\tableofcontents

\section{Introduction}\label{sec1}
Various universal events ~\cite{SupernovaSearchTeam:1998fmf,Planck:2015mvg,Planck:2015bpv} have demonstrated that improvements to general relativity (GR) are essential at a geometric scale. GR, grounded on Riemannian geometry, illustrates the Levi-Civita connection and defines the Ricci curvature $R$ as the essential characteristic of spacetime. This approach omits torsion and non-metricity from the geometry. An alternate approach to characterise GR and Riemannian geometry is teleparallel gravity ~\cite{aldrovandi2012teleparallel}. This theory suggests that the gravitational force originates from torsion $\mathcal{T}$ rather than curvature $R$ in Riemannian geometry. This idea was formally proposed by Einstein, who had earlier utilised teleparallel geometry in his efforts to formulate a unified field theory ~\cite{einstein2005riemann}. A different technique, referred to as non-metricity or symmetric teleparallel gravity, was developed by Nester ~\cite{nester1998symmetric}. This concept suggests that non-metricity $\mathcal{Q}$, which is not impacted by curvature and torsion, acts as the mediator of gravitational interactions. In contrast to General Relativity, where gravity is conveyed through the affine connection instead of the physical manifold, non-metricity is essential in this alternative approach. The Einstein pseudo-tensor, which transforms into a genuine tensor in the geometric framework, is associated with the non-metricity $\mathcal{Q}$. Jimenez et al.~\cite{BeltranJimenez:2017tkd} expanded symmetric teleparallel gravity to align with general relativity, obtaining $F(\mathcal{Q})$ gravity. Given the growing focus on extended gravity theories in modern cosmology, researchers are exploring alternate geometries to achieve a deeper understanding of late cosmic acceleration ~\cite{Ferraro:2008ey,Geng:2011aj,Cai:2015emx,Jarv:2015odu}. Furthermore, it is important to acknowledge that the metric theories may be expanded beyond Riemannian geometry, as shown in ~\cite{Conroy:2017yln}. Harko et al.~\cite{Harko:2018gxr} presented the modified $F(\mathcal{Q})$  gravity by utilizing the metric-affine formalism and an expansion of symmetric teleparallel gravity. Recent research on advanced theories of gravity emphasises the critical examination of the astrophysical and cosmological characteristics of $F(\mathcal{Q})$ gravity. In this way, Hohmann et al.~\cite{Hohmann:2018wxu} investigated the propagation velocity and possible polarizations of gravitational waves within the framework of Minkowski spacetime in $F(\mathcal{Q})$ gravity. Soudi et al.~\cite{ Soudi:2018dhv} defined the strong-field characteristics of this gravitational theory by the examination of gravitational wave polarizations. A multitude of studies has investigated  $F(\mathcal{Q})$ gravity across diverse contexts, including the analysis of observational data constraints for various parametrizations of  $F(\mathcal{Q})$ ~\cite{Lazkoz:2019sjl,Ayuso:2020dcu}, energy conditions ~\cite{Mandal:2020lyq}, cosmography  ~\cite{Mandal:2020buf}, bouncing scenarios ~\cite{Mandal:2021wer,bajardi2020bouncing}, black holes~\cite{DAmbrosio:2021zpm}, and the development of the growth index in matter perturbations, among other issues~\cite {Nguyen:2021cnb}. Furthermore, comprehensive analyses of the $F(\mathcal{Q})$ theory may be found in contemporary literature, including~\cite{Esposito:2021ect,Atayde:2021pgb,Kimura:2020mpk,Dimakis:2021gby}. It is crucial to highlight that $F(\mathcal{Q})$ gravity precisely accounts for the universe's fast expansion with statistical accuracy comparable to other established modified gravity theories. Therefore, altered gravity theories at an astrophysical scale must be taken in order to deal with this so-called degeneracy. Promising approaches for this purpose include direct monitoring of the galactic core ~\cite{Mann:2018xkm} and the increasing number of gravitational wave findings ~\cite{LIGOScientific:2016aoc}. For a comprehensive examination of this subject, the subsequent sources ~\cite{errehymy2022anisotropic,Maurya:2022cyv,Tangphati:2021tcy} include modern contributions across several settings. Stellar evolution describes the ongoing alterations in a star's interior composition from its starting point to its final extinction. During this transforming process, a star releases a significant quantity of energy as photons or neutrinos.
The fascinating properties of these celestial bodies have inspired many scholars to investigate their significance in astronomy and astrophysics. Gravitational collapse is a crucial topic in relativistic astrophysics, intimately linked to the formation of self-gravitating objects. A stable system initially sustains hydrostatic equilibrium until its gravitational forces continuously decrease its size, leading to gravitational collapse. The collapse yields many configurations, including white dwarfs, neutron stars, black holes and gravastars, called compact stars contingent upon the star's starting mass. The precise solution of a novel gravastar model, based on the method in~\cite{Mazur:2004fk}, is formulated inside the framework of general relativity, specifically by integrating the cloud of strings and quintessence as presented in~\cite{Javed:2023nls}.  The composition and features of compact stars have fascinated scientists, leading to inquiries into their many developmental phases and intrinsic properties within the field of gravitational physics. Multiple endeavours have been undertaken to examine the properties of stellar models influenced by diverse physical forces. In this context, the concept of anisotropy is essential for examining the structural processes of compact structures ~\cite{Herrera:1997plx}.
Bower and Liang ~\cite{bondi1992anisotropic} examined regionally anisotropic pressure distributions inside spherically symmetric matter, demonstrating that pressure anisotropy significantly influences the parameters defining the hydrostatic equilibrium of stellar systems.
Maurya and his associates~\cite{Maurya:2017uyk}. examined compact spherically symmetric systems characterized by anisotropic pressure profiles, providing substantial insights into the stability analysis of small objects. Karamakar and colleagues ~\cite{paul2015relativistic} investigated the influence of anisotropy in cool compact stars. A large number of scholars have exhibited a keen interest in examining the intrinsic and extrinsic characteristics of celestial bodies using gravitational elements.
The Karmarkar condition is a crucial framework that establishes a relationship between temporal and radial metric elements, hence improving the examination of compact objects. The Karmakar condition deals with the embedding of a four-dimensional Riemannian manifold into a higher-dimensional manifold. This approach is utilized to investigate the solutions of static spherical spacetime and to analyze the physically viable characteristics of anisotropic stars ~\cite {Maurya:2016ecj, Singh:2016nfw}. Bhar et al. ~\cite{Bhar:2017jbl} used this technique to investigate the stability requirements of anisotropic compact stars.\\
Uneven primary stresses or the so-called anisotropic fluid are expected to appear when the densities of compact objects are typically higher than the density of nuclear matter. This often indicates that these compact objects have two distinct types of pressures: the tangential pressure $(p_{t})$ and the radial pressure $(p_{r})$ \cite{Herrera:1997plx}. The radial pressure component $(p_{r})$ is not equal to the components in the transverse direction $(p_{t})$, resulting in the anisotropic condition. J.H. Jeans initially predicted this effect for self-gravitating objects in the Newtonian regime in 1922 \cite{Hueckstaedt:2006bt}.  Shortly after, Lemaitre\cite{luminet2015lemaitre} had also taken into account the local anisotropy effect in the context of GR and shown that the higher limits placed on the maximum value of the surface gravitational potential might be relaxed. Ruderman \cite{ruderman1972pulsars} provided an intriguing overview of more realistic stellar models and demonstrated that an anisotropic star is likely to have a matter density ($\rho> 10^{15} gm/cm^3$), where the nuclear interaction takes on a relativistic character. Additionally, it is evident from Herrera's recent explanation \cite{ Herrera:2020gdg} that even if the starting configuration had isotropic pressure, it would vanish and the system would become anisotropic owing to energy dissipation in stellar growth.  However, a system has no cause to lose its acquired anisotropy in the last stage of dynamic development if the matter distribution is already anisotropic. One of the main characteristics of the growth of massive stars is this dissipation, which is caused by the emission of massless particles like photons or neutrons.\\
Makalo et al. ~\cite{Makalo:2023anc} examined exact solutions for charged stars employing the Karmarkar condition and found that the relationship between metric components and charge leads to the formulation of a viable anisotropic model. Sharif and Gul ~\cite{gul2024physical} investigated physically feasible and stable anisotropic stellar models in the context of the Karmarkar framework.
The physical characteristics of stellar objects have been analyzed in several modified gravitational theories utilising the Karmarkar condition ~\cite{Abbas:2019hyp,Sharif:2022cjv,hassan2023decoupled}. Spherically symmetric solutions are examined extensively within the context of  $F({R})$ gravity~\cite{Multamaki:2006zb,delaCruz-Dombriz:2009pzc,Kobayashi:2008tq,Upadhye:2009kt,Manzoor:2022fjd}. The same symmetry is also examined inside the $F(\mathcal{T})$ framework ~\cite{Aftergood:2014wla,DeBenedictis:2016aze,lin2017spherical,Krssak:2015oua}. The dynamic characteristics of compact stars have been examined in several alternative theories of gravity, such as $F(\mathcal{G})$ ~\cite{Abbas:2014vka,malik2022bardeen}, $F(\mathcal{Q})$ ~\cite{Mandal:2021qhx,Errehymy:2022gws}, and $F(\mathcal{G,T})$ ~\cite{Shamir:2017rjz, AwaisSadiq:2022pwn}, among others. Numerous other investigations on compact stars exist within various modified theories and scenarios~\cite{Bhar:2017jbl,Malik:2024eui,Malik:2024boe, Malik:2024vjm, Malik:2024nfl,Malik:2024tcg,Ashraf:2024cww,Bhar:2020tah,Bhar:2021fzc,Bhar:2017ynp}.\\
Covariant $F(\mathcal{Q})$ gravity is used by Narawade et al. in~\cite{Narawade:2024pxb} to recreate a cosmological model in Connection-III and FLRW spacetime. Several observational datasets and the H(Z) Hubble parameter are used to extensively evaluate the dynamic behavior of the reconstructed model. This study presents strong evidence for cosmic acceleration, indicating that $F(\mathcal{Q})$ gravity can successfully replace the $\Lambda CDM$ model and makes a strong case for an alternate, non-cosmological constant justification of the Universe's present accelerating expansion. Using a dynamical system method, Lohakare et al.~\cite {Lohakare:2024oeu} investigated the cosmic stability of $F(\mathcal{Q,B})$ gravity, where  $F(\mathcal{Q})$  stands for the nonmetricity scalar and  $F(\mathcal{B})$  for the boundary term. According to this study, the $F(\mathcal{Q,B})$ model is a good substitute for the conventional $\Lambda CDM$ model as it provides a strong framework for understanding the dynamics of cosmic expansion and accurately depicts the observed acceleration of the Universe. In ~\cite {Alwan:2024lng} Alwan et al. investigate neutron stars with covariant $F(\mathcal{Q})$ gravity, assuming that the stars are spherically symmetric and static with ideal fluid matter. Numerical techniques employing quadratic, exponential, and logarithmic $F(\mathcal{Q})$ models are used to determine neutron star characteristics including mass, radius, and compactness. A cosmological model within the context of symmetric teleparallel gravity is presented by Narawade et al.in ~\cite{Narawade:2023rip}. They discover a violation of the strong energy requirement, which may offer important insights into the nature of dark energy. In ~\cite{Narawade:2023tnn}, the accelerating cosmological model's dynamical aspect was examined. Additionally, the model's evolutionary behaviour of density parameters for the matter-dominated, radiation-dominated, and dark energy phases is demonstrated within the framework of the $F(\mathcal{Q})$ gravity, a modified symmetric teleparallel gravity. In~\cite{Narawade:2022cgb}, a cosmological model of the Universe is presented within the framework of $F(\mathcal{Q})$ gravity, and it is shown that $F(\mathcal{Q})$ gravity offers an alternative to dark energy for explaining the current cosmic acceleration.

This work mostly focused on $F(\mathcal{Q})$ gravity. The modified gravity offers a clear formulation of classical GR, enhancing inertial gravitational interaction while disregarding the affine spacetime framework. The distinctive characteristic of $F(\mathcal{Q})$ gravity is its ability to elucidate the Universe's late-time acceleration without necessitating modifications to scalar fields~\cite {Akarsu:2010zm,BeltranJimenez:2019tme, Lazkoz:2019sjl}. The efficacy of the $F(\mathcal{Q})$ gravity configuration was demonstrated across various domains, involving wormhole geometry~\cite {Harko:2018gxr, Xu:2019sbp, DAmbrosio:2020nev, Maurya:2022vsn, Maurya:2023kjx, Banerjee:2021mqk, Kiroriwal:2023nul} and dynamical analysis~\cite{Lu:2019hra}, encompassing cosmological aspects~\cite{Mandal:2020buf}, and has been articulated in relation to $F(\mathcal{Q})$ gravity. The mathematical expression $F(\mathcal{Q})= \mathcal{Q} + m \mathcal{Q}^{n}$ has been employed to examine holographic dark energy~\cite{Shekh:2021ule}, demonstrating that the deceleration parameter transitions from negative to positive, signifying a shift in the universe via an initial phase of deceleration to the present phase of acceleration. Within the context of extended non-metric theories of gravity, Capozziello et al. ~\cite{Capozziello:2022tvv} investigated the potential for cosmic slow-roll inflation.

They especially discovered that the usual conformal structure is reinstated in a high-energy domain, but it is disrupted in the weak field limit, as shown by generic functions $F(\mathcal{Q})$, notably $F(\mathcal{Q}) = \alpha Q + \beta Q^{m}$. Despite widespread acceptance, the teleparallel formulation of General Relativity parallels both STGR and GR ~\cite{Heisenberg:2022mbo}. The main distinction is that gravity is assigned to non-metricity $(\mathcal{Q})$, whereas curvature and torsion are considered negligible in STGR. One may always use a coincident gauge that is consistent for the non-metricity and zero value of the torsion limitation \cite{BeltranJimenez:2017tkd}.  

This approach is distinctive since it removes the affine connections. Nonetheless, the formulation of metrics will vary across distinct coordinate systems if we persist with the coincident gauge~\cite{Lu:2019hra} and regard the metric as the sole essential component in $F(\mathcal{Q})$ theory. Although investigations into extending gravity theories remain still developing, the symmetric teleparallel, or $F(\mathcal{Q})$ gravity has garnered significant interest. Owing to the innovative characteristics of $F(\mathcal{Q})$ gravity, several investigations was undertaken to explore diverse facets of the universe, taking cosmology, energy demands, covariant formulation, spherically symmetric structures, and signatures that underscore the importance of gravity. The premise of the $F(\mathcal{Q})$ theory of gravity is that, in contrast to the $F(\mathcal{R})$ theory, which entails fourth-order differential equations, the field equations of $F(\mathcal{Q})$ gravity are second-order~\cite{Lin:2021uqa}. This theory enhances the theoretical framework for modeling compact stars by providing novel methodologies for analyzing their structure, stability, and observational characteristics. It offers a viable avenue for tackling difficulties related to extreme astrophysical circumstances and testing the boundaries of our comprehension of gravity, which is a notable consequence.\\

This study presents a novel anisotropic compact star model in $F(\mathcal{Q})$ gravity derived from the resolution of Karmarkar condition in static spherical spacetime. There exist two categories of solutions to the Karmarkar condition one is the interior Schwarzschild solution (1916) possesses an inaccurate causality condition and the other is cosmic. In our research, we use a particular kind of metric potential characterized by a hyperbolic function. This is feasible within the physical constraints of the astrophysical model. Developing solutions can be utilized to characterize a physically plausible distribution of astrophysical matter. We have conducted a comprehensive analysis of all physical characteristics and provide sample figures to substantiate our results. The structure of the article is as follows:\\Section~\ref{sec1}: Introduction\\
Section~\ref{sec2}: In this section, we have discussed the field equations and the Karmakar condition.\\
Section~\ref{sec3}: In this section, we explain the physical properties and characteristics of our proposed model.\\
Section~\ref{sec4}. This section contains the conclusion part.

\section{Field equations of $\mathcal{F}(\mathcal{Q})$ gravity and Karmarkar condition}\label{sec2}

\subsection{Field equations of $\mathcal{F}(\mathcal{Q})$ gravity }\label{sec2.1}
The Levi-Civita affine connection is crucial in explaining and analysing general relativity. 
When discussed inside a space-time manifold, it aligns with the metric. Various manifolds and affine connections may be utilized, along with diverse theories of gravity.~\cite{Harada:2020ikm, BeltranJimenez:2019tme}
.The Levi-Civita connection identifies nonmetricity $\mathcal{Q}$ and torsion $\mathcal{T}$  as important geometrical concepts alongside curvature. Alleviating these limitations permits the exploration of non-Riemannian geometric theories characterized by non-zero torsion, nonmetricity, and curvature. The $\mathcal{F}(\mathcal{Q})$ gravity model has been established under these circumstances, resulting in a gravitational Lagrangian that includes an undefined non-metricity ($\mathcal{Q}$). The generalization of $\mathcal{F}(\mathcal{Q})$ gravity is very consequential for the expanding of the Universe. Jimenez et al.~\cite{BeltranJimenez:2019tme} were the pioneers in introducing the term "symmetric teleparallel gravity," often referred to as $\mathcal{F}(\mathcal{Q})$ gravity. Based upon the research of ~\cite{Zhao:2021zab} within the framework of $\mathcal{F}(\mathcal{Q})$ gravity, a generic metric spacetime is analyzed, in which the metric tensor $g_{\gamma \Upsilon }$ and the connection coefficients $\Gamma^{\sigma}_{\gamma \Upsilon}$ are treated as distinct entities.The subsequent equation delineates the Nonmetricity of the aforementioned connection:\\
 \begin{eqnarray} \label{eq1}
\mathcal{Q}_{\psi \gamma \Upsilon}=\nabla_{\psi}g_{\gamma  \Upsilon} =\partial_{\psi}g_{\gamma  \Upsilon}-\Gamma^{\sigma}_{\psi \gamma }g_{\sigma \Upsilon}-\Gamma^{\sigma}_{\psi \Upsilon}g_{\gamma  \sigma}.
\end{eqnarray}
where the affine connection is:
\begin{eqnarray}\label{eq2}
\Gamma^{\sigma}_{\gamma  \Upsilon}=\{^{\sigma}{}_{\gamma  \Upsilon}\}+\zeta^{\sigma}_{\gamma  \Upsilon}+L^{\sigma}_{\gamma  \Upsilon},
\end{eqnarray}
where $\{^{\sigma}{}_{\gamma  \Upsilon}\}$ denotes the levi-Civita connection, $\zeta^{\sigma}_{\gamma  \Upsilon}$ denote the Contortion tensor and $L^{\sigma}_{\gamma  \Upsilon}$ denotes the disformation. $\{^{\sigma}{}_{\gamma  \Upsilon}\}$ is solved by using the metric potential $g_{\gamma  \Upsilon}$ :
\begin{eqnarray}\label{eq3}
\{^{\sigma}{}_{\gamma  \Upsilon}\} \equiv \frac{1}{2}g^{\sigma \Psi}(\partial_{\gamma}g_{\Psi \Upsilon}+\partial_{\Upsilon}g_{\Psi \Upsilon}-\partial_{\Psi}g_{\gamma  \Upsilon}),
\end{eqnarray}
 $\zeta^{\sigma}_{\gamma  \Upsilon}$ is written as:
\begin{eqnarray}\label{eq4}
\zeta^{\sigma}_{\gamma  \Upsilon} \equiv \frac{1}{2} \mathcal{T}^{\sigma}_{\hspace{0.1cm}} {\gamma  \Upsilon}+\mathcal{T}_{(\gamma \hspace{0.3cm} \Upsilon)}^{\hspace{0.2cm}\sigma}
\end{eqnarray}
and $L^{\sigma}_{\gamma  \Upsilon}$ is written  as:
\begin{eqnarray}\label{eq5}  
L^{\sigma}_{\gamma  \Upsilon} \equiv \frac{1}{2} \mathcal{Q}^{\sigma}_{\gamma  \Upsilon}-\mathcal{Q}_{( \gamma\hspace{0.3cm} \Upsilon)}^{\hspace{0.2cm}\sigma}.
\end{eqnarray}
The nonmetricity conjugate is characterized as follows:

\begin{eqnarray}\label{eq6}
   H^{\psi}_{\gamma  \Upsilon} = \frac{-1}{4}\mathcal{Q}^{\psi}_{\gamma  \Upsilon} + \frac{1}{2}\mathcal{Q}^{\psi}_{(\gamma  \Upsilon)} + \frac{1}{4}(\mathcal{Q}^{\psi} -\Tilde {\mathcal{Q}}^{\psi})g_{\gamma  \Upsilon}-\frac{1}{4}\delta^{\psi}_{(\gamma  \Upsilon)}.
   \end{eqnarray}
The independent derivations of the aforementioned equations are outlined below:
\begin{eqnarray}\label{eq 7}
\mathcal{Q}_{\psi} \equiv \mathcal{Q}_{\psi \gamma}^{\hspace{0.2cm} \gamma}\hspace{0.6cm}\Tilde{\mathcal{Q}}_{\psi} \equiv \mathcal{Q}^{\gamma}_{\hspace{0.1cm}\psi\hspace{0.1cm}\gamma},
\end{eqnarray}
now, the $\mathcal{Q}$ is given as: 
\begin{eqnarray}\label{eq8}
\mathcal{Q}= -\hspace{0.1cm} \mathcal{Q}_{\psi  \Upsilon \gamma}F^{\psi \gamma  \Upsilon}.
\end{eqnarray}
The $\mathcal{F}(\mathcal{Q})$ gravity is characterized by a specific action via Lagrangian distribution:
\begin{eqnarray}\label{eq9}
S = \int \sqrt{-g}d^{4}x\Bigg[\frac{1}{2}\mathcal{F}(\mathcal{Q})+\sigma^{\Psi \gamma  \Upsilon}_{\psi}\mathcal{R}^{\psi}_{\Psi \gamma  \Upsilon} +\sigma^{\gamma  \Upsilon}_{\psi} \mathcal{T}^{\psi}_{\gamma  \Upsilon}+\mathcal{L}_{n}\Bigg].
\end{eqnarray}
In Eq.~(\ref{eq9}), $g$ signifies the determinant metric, $\mathcal{F}(\mathcal{Q})$ indicates the nonmetricity scalar $\mathcal{Q}$, and $\mathcal{L}_{n}$ specifies the Lagrange density, and $\sigma_{\psi}^{\Psi \gamma  \Upsilon}$ represents the Lagrange multipliers.
The field equations pertaining to the metric below have been derived using the action provided in Eq.~(\ref{eq9}) as: 
\begin{eqnarray}\label{eq10}
\mathcal{T}_{\gamma  \Upsilon} =\frac{2}{\sqrt{-g}}\Delta_{\psi}(\sqrt{-g} \mathcal{F}_{\mathcal{Q}}H^{\psi}_{\gamma  \Upsilon})\nonumber\\&&\hspace{-4.1cm} + \frac{1}{2}g_{\gamma  \Upsilon} \mathcal{F}+\mathcal{F}_{\mathcal{Q}}(H_{\gamma \psi \Psi} \mathcal{Q}_{\Upsilon}^{\psi \Psi}-2 \mathcal{Q}_{\psi \Psi \gamma}H^{\psi \Psi}_{\Upsilon}).
\end{eqnarray}

where, $\mathcal{F}_{\mathcal{Q}}=\frac{\partial\mathcal{F}}{\partial {\mathcal{Q}}}$, and The energy-momentum tensor (EMT) can be described as: 
\begin{eqnarray}\label{eq11}
\mathcal{T}_{\gamma  \Upsilon} \equiv \frac{2}{\sqrt{g}} \frac{\delta(\sqrt{-g})\mathcal{L}_{n}}{\delta g^{\gamma  \Upsilon}}.
\end{eqnarray}
We get the below system upon modifying equation Eq.~(\ref{eq9}) related to affine connection:
\begin{eqnarray}\label{eq12}
\nabla_{\rho}\sigma_{\psi}^{\Upsilon \gamma \rho}+\sigma_{\psi}^{\gamma  \Upsilon}=\sqrt{-g \mathcal{F}_{\mathcal{Q}}}H_{\gamma  \Upsilon}^{\psi}+J_{\psi}^{\gamma  \Upsilon}.
\end{eqnarray}
The hypermomentum tensor density is given as:
\begin{eqnarray}\label{13}
J_{\psi}^{\gamma  \Upsilon}= \frac{-1}{2} \frac{\delta \mathcal{L}_{n}}{\delta T_{\gamma  \Upsilon}^{\psi}}.
\end{eqnarray}

To simplify equation Eq.~(\ref{eq12}) to the following equation, we use the anti-symmetry property of s and t in the lagrangian multiplier coefficients:
\begin{eqnarray}\label{eq14}
\nabla_{\gamma}\nabla_{\Upsilon}(\sqrt{-g}\mathcal{F}_{\mathcal{Q}}H^{\gamma  \Upsilon}_{\hspace{0.3 cm}\psi}+ J_{\psi}^{\hspace{0.3 cm} \gamma  \Upsilon})= 0.
\end{eqnarray}

if we consider$ \nabla_{\gamma}\nabla_{\Upsilon}H_{\psi}^{\gamma  \Upsilon}=0$ we get :
\begin{eqnarray}\label{eq15}
\nabla_{\gamma}\nabla_{\Upsilon}(\sqrt{-g}\mathcal{F}_{\mathcal{Q}}H^{\gamma  \Upsilon}_{\hspace{0.3 cm}\psi})= 0.
\end{eqnarray}
 In absence of torsion or curvature, the affine connection takes the form: 
\begin{eqnarray}\label{eq16}
\Gamma^{\psi }_{\gamma  \Upsilon}=\Bigg( \frac{\partial{x}^{a}}{\partial{\eta}^\sigma}\Bigg)\partial_{\gamma}\partial{\Upsilon}\eta^\sigma.
\end{eqnarray}
When we examine a certain coordinate choice recognized as a coincident gauge, for which $\Gamma^{\psi}_{\gamma  \Upsilon}= 0$. After that, the nonmetricity becomes:

\begin{eqnarray}\label{eq17}
\mathcal{Q}_{\psi \gamma  \Upsilon}=\partial_{\psi}g_{\gamma  \Upsilon}.
\end{eqnarray}
The metric for a generic static and spherically symmetric (SS) spacetime  is articulated as: 
\begin{eqnarray}\label{eq18}
ds^{2}=-e^{\Upsilon(r)}dt^{2}+ e^{\sigma(r)}dr^{2}+r^{2}dw^{2}.
\end{eqnarray}

In the above equation $dw^{2}\equiv d\theta^{2}+sin^{2}{\theta} d{\phi}^{2}$. Equation ~(\ref{eq19}) is derived by substituting Equation ~(\ref{eq18}) into Equation ~(\ref{eq8}), then  $\mathcal{Q}$  is expressed in terms of r,
\begin{eqnarray}\label{eq19}
\mathcal{Q}(r) = -\frac{2e^{-\sigma}}{r}\Bigg(\Upsilon^{\prime}+\frac{1}{r}\Bigg),
\end{eqnarray}
where the derivative with respect to r is denoted by prime($\prime$) and  $\mathcal{T}_{\gamma  \Upsilon}$ is given as:

\begin{eqnarray}\label{20}
\mathcal{T}_{\gamma  \Upsilon}=(\rho+p_{t})u_{\gamma}u_{\Upsilon}+p_{t}g_{\gamma  \Upsilon}+(p_{r}-p_{t})v_{\gamma}v_{\Upsilon},
\end{eqnarray}
where, $u_{\gamma}$ denotes four-velocity, and $v_{\gamma}$ denotes the unitary space-like vector in the radial direction that fulfils $u^{\gamma} u_{\gamma}=-1,v^{\gamma} v_{\gamma}=1,u^{\gamma} u_{\gamma}=-1,u^{\gamma}v_{\Upsilon}=0$. Energy density is represented by $\rho(r)$, radial pressure in the direction of $v_{\gamma}$ is represented by $p_{r}(r)$, and tangential pressure is represented by the orthogonal. $p_{t}(r)$ represents $(v_{\gamma})$.In the equations of motion Eq.~(\ref{eq10}) for anisotropic fluid Eq.~(\ref{eq12}), each of the independent variables are expressed as follows: 
\begin{eqnarray}\label{eq21}
&&\hspace{-0.5cm} \rho=-\frac{\mathcal{F}}{2}+\mathcal{F}_{\mathcal{Q}}\Bigg[\mathcal{Q}+\frac{1}{r^{2}}+\frac{e^{-\sigma}}{r}(\Upsilon^{\prime}+\sigma^{\prime})\Bigg],\\
&& \hspace{-0.6cm} p _{r} = \frac{\mathcal{F}}{2}-\mathcal{F}_{\mathcal{Q}}\Bigg[\mathcal{Q}+\frac{1}{r^{2}}\Bigg]\label{eq22},\\
&& \hspace{-0.6cm} p_{t} = \frac{\mathcal{F}}{2}-\mathcal{F}_{\mathcal{Q}}\frac{\mathcal{Q}}{2} -\mathcal{F}_{\mathcal{Q}}e^{-\sigma}\frac{\mathcal{Q}}{2}\nonumber\\&&\hspace{-0.2cm} -\mathcal{F}_{\mathcal{Q}}e^{-\sigma}\Bigg[\frac{\Upsilon^{\prime\prime}}{2}+(\frac{\Upsilon^{\prime}}{4}+\frac{1}{2r})(\Upsilon^{\prime}-\sigma^{\prime})\Bigg],\label{eq23}\\
&& \hspace{-0.6cm} 0 = \frac{\cot\theta}{2}\mathcal{Q}^{\prime}\mathcal{F}_{\mathcal{Q} \mathcal{Q}}\label{eq24}.
\end{eqnarray}
With the assumption of a zero affine connection in the coordinate system, the gravity equation $\mathcal{F}(\mathcal{Q})$ can be stated as follows: 
\begin{eqnarray}\label{eq25}
 \frac{\cot\theta}{2}\mathcal{Q}^{\prime}\mathcal{F}_{\mathcal{Q} \mathcal{Q}}=0.
\end{eqnarray}
Combining equation (\ref{eq10}) and motion equations yields $\mathcal{F}_{\mathcal{Q} \mathcal{Q}} = 0$. This implies that the function $\mathcal{F}(\mathcal{Q})$ have to be linear. Selecting nonlinear form of $\mathcal{F}(\mathcal{Q})$ may not satisfy field equations and its solutions, particularly when $\mathcal{F}(\mathcal{Q}=\mathcal{Q}^{2})$ is used. The nonlinear function of $\mathcal{F}(\mathcal{Q})$ gravity requires an extensive SS metric for coincident gauges in order to be analyzed and resolved~ (see~\cite{Zhao:2021zab}). In this study, we employed a linear form of $\mathcal{F}_{\mathcal{Q} \mathcal{Q}} = 0$ and determined that the SS coordinate system (\ref{eq18}) aligns well with  $\Gamma^{\gamma}_{\gamma \Upsilon}=0$. 
In order to develop a compact stellar model that is realistic, it is necessary to identify a appropriate function of $\mathcal{F}(\mathcal{Q})$. The choice of a function that asserts $\mathcal{F}_{\mathcal{Q} \mathcal{Q}} = 0$ is essential to align the our metric with the exterior Schwarzschild (Anti-) de Sitter metric at the external boundary, according to ~\cite{Wang:2021zaz}. The function $\mathcal{F}(\mathcal{Q})$ is then employed to ascertain the field equations  that can be defined as follows:
\begin{eqnarray}\label{eq26}
\mathcal{F}_{\mathcal{Q} \mathcal{Q}}=0\Rightarrow \mathcal{F}(\mathcal{Q})= \delta_{1}\mathcal{Q}+\delta_{2}.
\end{eqnarray}
where, $\delta_{1}$ and $\delta_{2}$ are integration constants.  Using equations  Eq.~(\ref{eq25}) and  Eq.~(\ref{eq26}), the field equations for $\mathcal{F}(\mathcal{Q})$ gravity can be written as:
\begin{eqnarray}\label{eq27}
&&\hspace{-1.0cm} \rho= \frac{1}{16 \pi r^{2}}\Bigg[2\delta_{1}+2e^{-\sigma}\delta_{1}(r{\sigma}^{\prime}-1)-r^{2}\delta_{2}\Bigg],\\
&&\hspace{-1.0cm} p_{r}=\frac{1}{16 \pi r^{2}}\Bigg[-2\delta_{1}+2e^{-\sigma}\delta_{1}(r \Upsilon^{\prime}+1)+r^{2}\delta_{2}\Bigg],\label{eq28}\\
&&\hspace{-1.0cm} p_{t}=\frac{e^{-\sigma}}{ 32 \pi r}\Bigg[2e^{\sigma}r\delta_{2}+\delta_{1}(2+r\Upsilon^{\prime})(\Upsilon^{\prime}-\sigma^{\prime})+2r\delta_{1}\Upsilon^{\prime\prime}\Bigg].\label{eq29}
\end{eqnarray}
In Equations (\ref{eq27}--\ref{eq29}) The prime symbol $(\prime)$ signifies differentiation with respect to the radial coordinate \( r \), whereas \( \rho \), \( p_{r} \), \( p_{t} \), \( \sigma \), and \( \Upsilon \) are variables that we will resolve to get the desired outcome. For static and SS fluid systems, the generic formulation for the anisotropic component, written as $\Delta=(p_{t}-p_{r})$, is articulated as follows:  
\begin{equation}\label{eq30}
\Delta(r)=\frac{\delta_{1}}{8 \pi}\Bigg[e^{-\sigma}\Bigg(\frac{\Upsilon^{\prime\prime}}{2}-\frac{\sigma^{\prime}\Upsilon^{\prime}}{4}+\frac{(\Upsilon^{\prime})^{2}}{4}-\frac{\Upsilon^{\prime}+\sigma^{\prime}}{2r}-\frac{1}{r^{2}}\Bigg)+\frac{1}{r^{2}}\Bigg].
\end{equation}

\subsection{Karmarkar Condition}\label{sec2.2}
The SS line element (\ref{eq18}) may continually be embedded in six-dimensional flat spacetime, signifying its classification as belonging to embedding class two in general. On the contrary, the SS metric can be included in five-dimensional flat spacetime, provided it fulfils the Karmarkar criterion ~\cite{ karmarkar1948gravitational}. Then It indicates the spacetime of embedding class one. It is a necessary and sufficient condition for SS spacetime to belong to class one. The Karmarkar condition is expressed through the curvature components as follows:
\begin{equation}\label{eq31}
R_{1414}=\frac{R_{1212}R_{3434}-R_{1224}R_{1334}}{R_{2323}},
\end{equation}
with $R_{2323} \neq 0$ ~\cite{pandey1982insufficiency}. The nonzero components of the Riemann curvature tensor \( R_{hijk} \) for the metric (\ref{eq18}) are defined as follows:\\\\
$R_{2323} = \frac{\sin^{2}\theta(e^{\sigma}-1)r^{2}}{e^{-\sigma}}$,\quad $R_{1212}=\frac{\sigma^{\prime}r}{2}$,\\\\
$R_{2424}=-\frac{\Upsilon^{\prime}re^{\Upsilon-\sigma}}{4}$,\quad $R_{1334}=R_{1224}sin^{2}\theta =0$,\\\\ $R_{1414}=-\frac{1}{4e^{\Upsilon}}[2\Upsilon^{\prime \prime}+(\Upsilon^{\prime})^{2}-\sigma^{\prime}\Upsilon^{\prime}]$,\quad $R_{3434}=\sin^{2}\theta R_{2424}.$\\\\
By substituting these components of $R_{hijk}$ in  Eq.~(\ref{eq31}), we obtain the subsequent differential equation:
\begin{equation}\label{eq32}
\frac{2\Upsilon^{\prime \prime}+(\Upsilon^{\prime})^{2}}{\Upsilon^{\prime}}=\frac{\sigma^{\prime}e^{\sigma}}{(e^{\sigma}-1)}.
\end{equation}
When solving the differential  Eq.~(\ref{eq32}), we get the metric potential,
\begin{equation}\label{eq33}
e^{\Upsilon}=\Bigg[A_{1}+B_{1}\int\sqrt{(e^{\sigma(r)}-1)dr}\Bigg]^{2},
\end{equation}
where $A_{1}$ and $B_{1}$ are nonzero arbitrary constants of integration. Additionally, we choose the metric potential $e^{\sigma}$ as 
\begin{equation}\label{eq34}
e^{\sigma }=b^2 r^2 \text{csch}\left(f r^2+h\right) \text{csch}\left(f r^2+h\right)+1,
\end{equation}
where $b \neq 0$, $f \neq 0$ or $h \neq 0$ ~\cite{Prasad:2019epe}. If b = f = h = 0, then the reduced spacetime does not constitute a class one spacetime~\cite{pandey1982insufficiency}. In 2017, Bhar proposed that for all well-behaved models, the metric function $e^{\sigma}$ must be monotonically rising and fulfil the constraint $e^{\sigma}(0)=1$ and $((e^{\sigma})')_{r} = 0.$ It is noted that our metric function $e^{\sigma}$ is an increasing function and satisfy the aforementioned conditions. This indicates that $e^{\sigma}$, as defined by Eq.~(\ref{eq34}), is physically acceptable.
Substituting the value of $e^{\sigma}$ from Eq.~(\ref{eq34}) in Eq.~(\ref{eq33}) we ge the value of $e^{\Upsilon}$ as 
\begin{equation}\label{eq35}
e^{\Upsilon}=\Bigg[A_{1}+\frac{B_{1}b}{2 f}\log \Bigg(\frac{e^{(fr^{2}+h)}-1}{e^{(f r^{2}+h)}+1}\Bigg)\Bigg]^{2}.
\end{equation}
Using the value of $e^{\sigma}$ and $e^{\Upsilon}$ from Eq.~(\ref{eq34}) and Eq.~(\ref{eq35}) in  Eqs.~(\ref{eq27}--\ref{eq30}), we get the expressions for energy density ($\rho$), radial pressure ($p_{r}$), tangential pressure ($p_{t}$), and the anisotropic factor ($\Delta$) as follows:
\begin{eqnarray}\label{eq36}
8 \pi \rho =-\frac{b^4 r^2 \left(\delta _2 r^2-2 \delta _1\right) \text{csch}^4\left(f r^2+h\right)+\delta _2+\phi _1}{2  \left(b^2 r^2 \text{csch}^2\left(f r^2+h\right)+1\right)^2},\\
&&\hspace{-8.5cm} 8 \pi p_{r}=\frac{8 B_1 b \delta _1 f r^2 \log (e) e^{f r^2+h}+\delta _1 \phi _2}{ r^2 \phi _3 \left(e^{f r^2+h}-1\right) \left(e^{fr^2+h}+1\right)}+ \frac{\delta _2}{2}-\frac{\delta _1}{ r^2},\\ \label{eq37}
&&\hspace{-8.5cm} 8\pi p_{t}=\frac{\delta _2 \phi _5 \left(b^4 r^4 \text{csch}^4\left(f r^2+h\right)+1\right)+2 b^2 \phi _6+\phi _4}{2 \phi _7 \left(e^{f r^2+h}-1\right)^2 \left(e^{f r^2+h}+1\right)^2},\label{eq38}\\
&&\hspace{-8.5cm} 8\pi \Delta =\frac{\delta _2 \phi _5 \left(b^4 r^4 \text{csch}^4\left(f r^2+h\right)+1\right)+2 b^2 \phi _6+\phi _4}{2 \phi _7 \left(e^{f r^2+u}-1\right)^2 \left(e^{f r^2+h}+1\right)^2}\\
\nonumber&&\hspace{-7.5cm}-\frac{8 B_1 b \delta _1 f r^2 \log (e) e^{f r^2+h}+\delta _1 \phi _2}{r^2 \phi _3 \left(e^{f r^2+h}-1\right) \left(e^{f r^2+h}+1\right)}-\frac{\delta _2}{2}+\frac{\delta _1}{r^2}.\label{eq39}
\end{eqnarray} 
The value of $\phi_{1}$,$\phi_{2}$,$\phi_{3}$,$\phi_{4}$,$\phi_{5}$,$\phi_{6}$ and $\phi_{7}$ are given in Appendix.

\begin{table*}[!htp]
\centering
\begin{tabular}{cccccccccc}
      \hline
    Compact star&\hspace{0.5cm}$M(M_{0})$&\hspace{0.5cm}$R(km)$&\hspace{0.5cm}$\delta_{1}$&\hspace{0.5cm}$b( km^{-1})$ & \hspace{0.5cm}$f( km^{-1})$ &\hspace{0.5cm} $h$&\hspace{0.5cm}$\frac{M}{R}$ &\hspace{0.5cm}$A_{1}$&\hspace{0.5cm}$B_{1}$\\
    \hline
  Cen X-3 &\hspace{0.5cm}1.48&\hspace{0.5cm}11.6&\hspace{0.5cm}0.2&\hspace{0.5cm}0.107&\hspace{0.5cm}-0.0011&\hspace{0.5cm}1.39&\hspace{0.5cm}0.128291379&\hspace{0.5cm}0.0269800&\hspace{0.5cm}0.0263286\\
    \hline
   Cen X-3&\hspace{0.5cm}1.49&\hspace{0.5cm}11.6&\hspace{0.5cm}0.4&\hspace{0.5cm}0.107 &\hspace{0.5cm}-0.0011&\hspace{0.5cm}1.39&\hspace{0.5cm}0.128481897&\hspace{0.5cm}0.0235791&\hspace{0.5cm}0.0264463 \\
    \hline
    Cen X-3&\hspace{0.5cm}1.49&\hspace{0.5cm}11.6&\hspace{0.5cm}0.6&\hspace{0.5cm}0.107&\hspace{0.5cm}-0.0011&\hspace{0.5cm}1.39&\hspace{0.5cm}0.128544828&\hspace{0.5cm}0.0224455&\hspace{0.5cm}0.0264855\\
    \hline 
   Cen X-3&\hspace{0.5cm}1.49&\hspace{0.5cm}11.6&\hspace{0.5cm}0.8&\hspace{0.5cm}0.107&\hspace{0.5cm}-0.0011&\hspace{0.5cm}1.39&\hspace{0.5cm}0.128576724&\hspace{0.5cm}0.0218787&\hspace{0.5cm}0.0265051 \\
    \hline
    Cen X-3&\hspace{0.5cm}1.49&\hspace{0.5cm}11.6&\hspace{0.5cm}1.0&\hspace{0.5cm}0.107&\hspace{0.5cm}-0.0011&\hspace{0.5cm}1.39&\hspace{0.5cm}0.128671552&\hspace{0.5cm}0.0201782&\hspace{0.5cm}0.026564 \\ 
    \hline
    \end{tabular}
    \caption{The numerical value of Mass $M(M_{0})$, Radius $R(km)$, $b$, $f$, $h$, $A_{1}$ and $B_{1}$ for different values of $\delta_{1}$}
    \label{Table1}
\end{table*}

\begin{table*}[!htp]
\centering
\begin{tabular}{cc}
    Expression\hspace{8.2cm}&Remarks\\
    \hline
 $ \rho, p_{r}, p_{t}$ &$>0$, acceptable\\
    \hline
   $\frac{p_{rc}}{\rho_{c}}$&\hspace{-1.7cm}$\leq0$ \\
    \hline
    $\Delta$&$>0$, acceptable\\
    \hline 
   $\frac{d\rho}{dr}, \frac{dp_{r}}{dr}, \frac{dp_{t}}{dr}$ &$<0$, acceptable\\
    \hline
   $F_{a}, F_{h}, F_{g}$ & balanced \\ 
    \hline
    $\rho \pm p_{r}$&$>0$, acceptable \\
    \hline
    $\rho \pm p_{t}$&$>0$, acceptable \\
    \hline
    $\rho +p_{r}+2p_{t}$&$>0$, acceptable \\
    \hline
   $m(r)$&$>0,acceptable$ \\
    \hline
     $u(r)$&\hspace{1.3cm} $0<u(r)<\frac{8}{9}$, acceptable \\
     \hline
     $Z_{S}$&\hspace{1.2cm}$0<Z_{S}<5$, acceptable \\
     \hline
$\omega_{r}, \omega_{t}$ &\hspace{1.6cm}$0<\omega_{r}, \omega_{t}<1$, acceptable\\
\hline
$V^{2}_{r}, V^{2}_{t}$&\hspace{1.8cm}$0<V^{2}_{r}, V^{2}_{t}<1$, acceptable \\
\hline
$V^{2}_{t}-V^{2}_{r}$&\hspace{2.0cm}$-1<|V^{2}_{t}-V^{2}_{r}|<1$, acceptable \\
\hline
$\Gamma$&$>\frac{4}{3}$, acceptable \\
\hline
 \end{tabular}
    \caption{The table shows the acceptability of all parameters of the proposed compact star}
    \label{Table2}
\end{table*}

\section{Physical properties and characteristics of the model }\label{sec3}
In this section, we describe the physical properties and characteristics of an anisotropic compact star.

\subsection{Metric elements}\label{sec3.1}

\begin{figure*}[!htp]
    \centering
    \includegraphics[height=7cm,width=7.5cm]{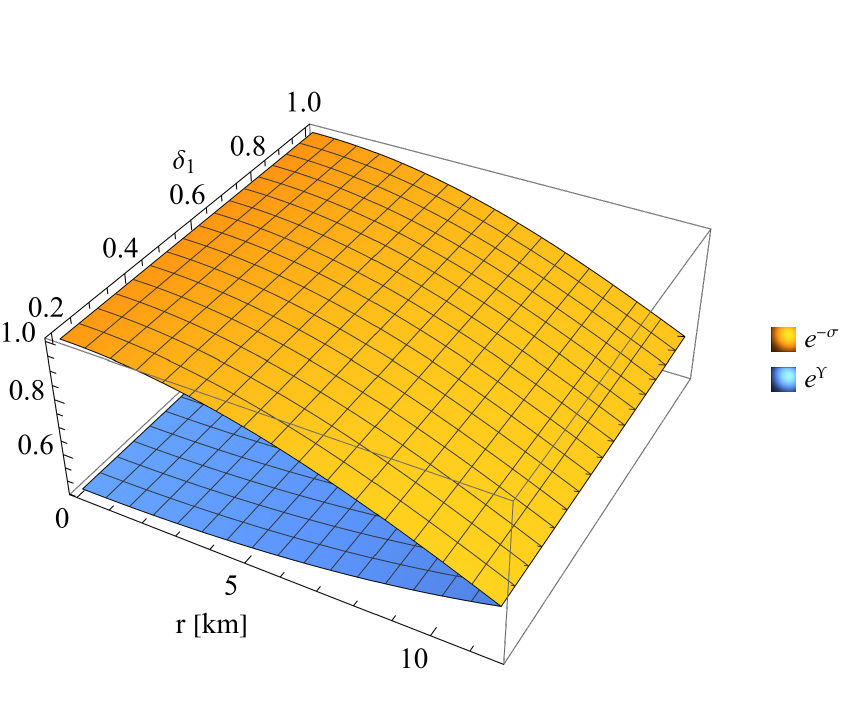}
    \caption{Graphical representation of metric elements $e^{-\sigma} $ and $ e^{\Upsilon}$ for fixed value of $\delta_{2}=0.00001$ versus   $\delta_{1}$ and radius r of the proposed compact star.}
    \label{figure 1}
\end{figure*}

Anisotropic compact stars require finite and nonsingular metric elements to produce a physically plausible model. 
According to our solution, $e^{\sigma}$ at $r=0$ equals 1 and $e^{\Upsilon}$ at r=0 equals 0.4583. At r=0, we see a positive value of $e^{\sigma}$ and $e^{\Upsilon}$, showing that the metric components are consistent at the core of the object. The precise behavior of the metric components is displayed in Fig.~\ref{figure 1}. The graphs of the components $e^{-\sigma}$ and $e^{\Upsilon}$ are bounded and nonsingular within the stellar configuration. The graph of $e^{-\sigma}$ exhibits a monotonically decreasing trend near the surface, whereas the graph of $e^{\Upsilon}$ demonstrates a monotonically increasing trend approaching the surface. The graph intersects near the border of the compact star; given these qualities of the metric components, we may infer that they are capable of generating a model for the anisotropic stellar model.

\subsection{Seamless Matching of Interior and Exterior spacetime}\label{sec3.2}
This section presents the boundary condition necessary for resolving the Einstein field equations in $\mathcal{F}(\mathcal{Q})$ gravity. To do this, we compare the solutions of the outer and inner spacetimes. Recent studies, such as~\cite{Wang:2021zaz}, suggest that the viable solution in $\mathcal{F}(\mathcal{Q})$ gravity may be resolved using the Schwarzschild (Anti-) de Sitter solution. The result is presented as follows:  

\begin{eqnarray}\label{eq40}
ds^{2}=-\Bigg(1-\frac{2M}{r}-\frac{\Lambda}{3}r^{2} \Bigg)dt^{2} + \frac{dr^{2}} {\Bigg(1-\frac{2M}{r}-\frac{\Lambda}{3}r^{2}\Bigg)} \nonumber\\&&\hspace{-7.3cm}+\hspace{0.1cm} r^{2}(d\theta^{2}+\sin^{2}\theta d\phi^{2}).
\end{eqnarray}
By setting \( r = R \), the Schwarzschild (Anti-)de Sitter metric has been solved at the boundary in accordance with the criteria established by Darmois-Israel in their study ~\cite{Israel:1966rt} as: 

\begin{eqnarray}\label{eq41}
&& \hspace{-0.5cm}\Bigg(1-\frac{2M}{r}-\frac{\Lambda}{3}r^{2}\Bigg)=e^{\Upsilon(R)},\\
&& \hspace{-0.5cm}
\Bigg(1-\frac{2M}{r}-\frac{\Lambda}{3}r^{2}\Bigg)=e^{-\sigma(R)}\label{eq42},\\
&& \hspace{1.5cm}p_{r}(R)=0.\label{eq43}
\end{eqnarray}\\
To ensure continuity of the first derivative:
\begin{equation}\label{eq44}
\Bigg(\frac{\partial w^{-}_{ss}}{\partial r}\Bigg)_{r=R}= \quad \Bigg(\frac{\partial w^{+}_{ss}}{\partial r}\Bigg)_{r=R}
\end{equation}
and
\begin{equation}\label{eq45}
 w^{-}_{ss} = e^{\Upsilon}, \quad w^{+}_{ss} = \Bigg(1-\frac{2M}{r}-\frac{\Lambda}{3}r^{2}\Bigg).
\end{equation}
Now from Eq.~(\ref{eq44}) and Eq.~(\ref{eq45})  we get,
\begin{equation}\label{eq46}
\Upsilon^{\prime}(R) = \frac{6M-2\Lambda R^{3}}{3 R^{2}-6MR-\Lambda R^{4}}
\end{equation}
By solving equation Eq.~(\ref{eq28}) and  Eq.~(\ref{eq43}) weget,
\begin{equation}\label{eq47}
\Upsilon^{\prime}(R) = \frac{6M-2\Lambda R^{3}}{3 R^{2}-6MR-\Lambda R^{4}}
\end{equation}
The analysis from Eq.~(\ref{eq46}) and  Eq.~(\ref{eq47}) demonstrates the continuity of the first derivative at $r = R.$ The symbols $M$ and $\Lambda$ represent the mass and the cosmological constant, respectively.

The cosmological constant $\Lambda$ is dependent on the constants $\delta_{1}$ and $\delta_{2}$, represented by the equation $\Lambda =\frac{\delta_{2}}{2\delta_{1}}$. In this context, the evidence for the cosmological constant is nonexistent; hence, its value does not influence the present star model. The cosmological constant often has a substantial value that affects the mathematical formulation of the issue. The cosmological constant, currently assessed as zero in this research, is projected to be $10^{-46}km^{-2}$ according to numerical evidence. In this project, $\delta_{2}$ has been established as 0.00001 throughout the duration of the problem. The parameter $\delta_{1}$ significantly influences the suggested model; an increase in $\delta_{1}$ results in corresponding increases in the mathematical values and graphical behaviors of physical parameters such as density, pressure, adiabatic index, mass, compactness, and sound velocity. Simultaneously, the center density and central pressure of the stellar model also increase.\\

 Equations (\ref{eq41}--\ref{eq43}) represent boundary conditions, which facilitated the calculation of the constants \( A_{1} \) and \( B_{1} \) as, 

\begin{eqnarray}\label{eq48}
&& \hspace{-0.9cm} A_{1}=\frac{B_1 b \left(-\phi _{10} \left(\delta _2+\phi _{11}\right)-16 \delta _1 f \log (e) e^{f R^2+h}\right)}{2 f \phi _{12} \left(e^{2 \left(f R^2+h\right)}-1\right)},\\
&& \hspace{-0.9cm}
B_1=-\frac{\phi _{13} e^{-f R^2-h} \left(e^{2 \left(f R^2+h\right)}-1\right)}{8 b \delta _1 \log (e)}\label{eq49}. 
\end{eqnarray}
The value of $\phi_{10}$,$\phi_{11}$,$\phi_{12}$, and $\phi_{13}$ are given in appendix

\subsection{Behaviour of density}\label{sec3.3}
For a well-behaved model of an anisotropic compact star, the density value is maximum at the centre and decreases as the radius increases towards the surface. In our model, the density value is optimum at the centre and decreases 
 as $r$ rises as shown in Fig.~\ref{figure 2}. Therefore our model is well-behaved. 

\subsection{Behaviour of radial and tangential pressure}\label{sec3.4}.
Here we analyse the graphical behaviour of radial pressure ($p_{r}$) and tangential pressure ($p_{t}$). Figure~\ref{figure 2} represents the behaviour of radial and tangential pressure, both pressures attain the Optimum value at the centre(r=0) and decrease monotonically as the radius increases towards the surface. Tangential pressure is always positive throughout the compact star, but radial pressure diminishes toward the border.

\subsection{Anisotropy factor and anisotropic force}\label{sec3.5}
\begin{figure*}[!htp]
    \centering
    \includegraphics[height=7cm,width=7.5cm]{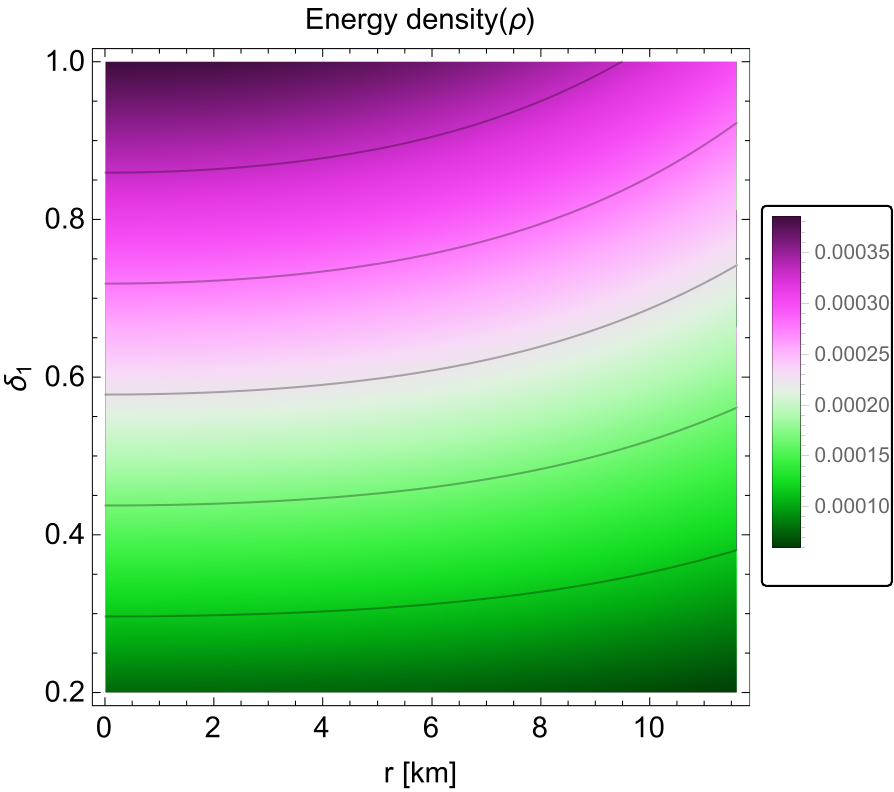}~~~~~~~~~\includegraphics[height=7cm,width=7.5cm]{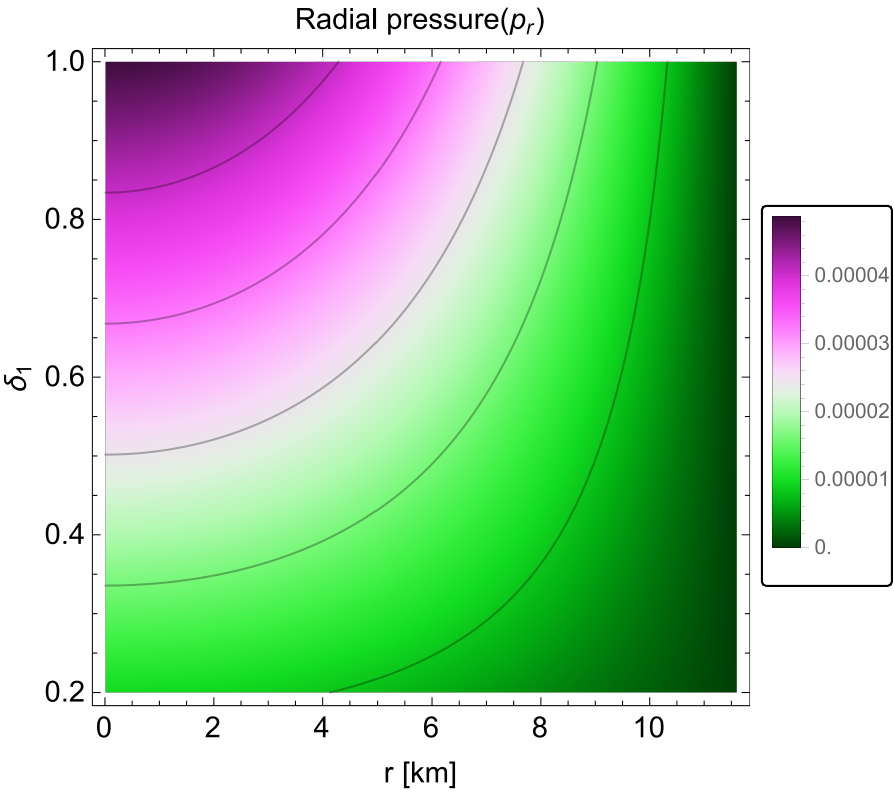}\\~~~~~\includegraphics[height=7cm,width=7.5cm]{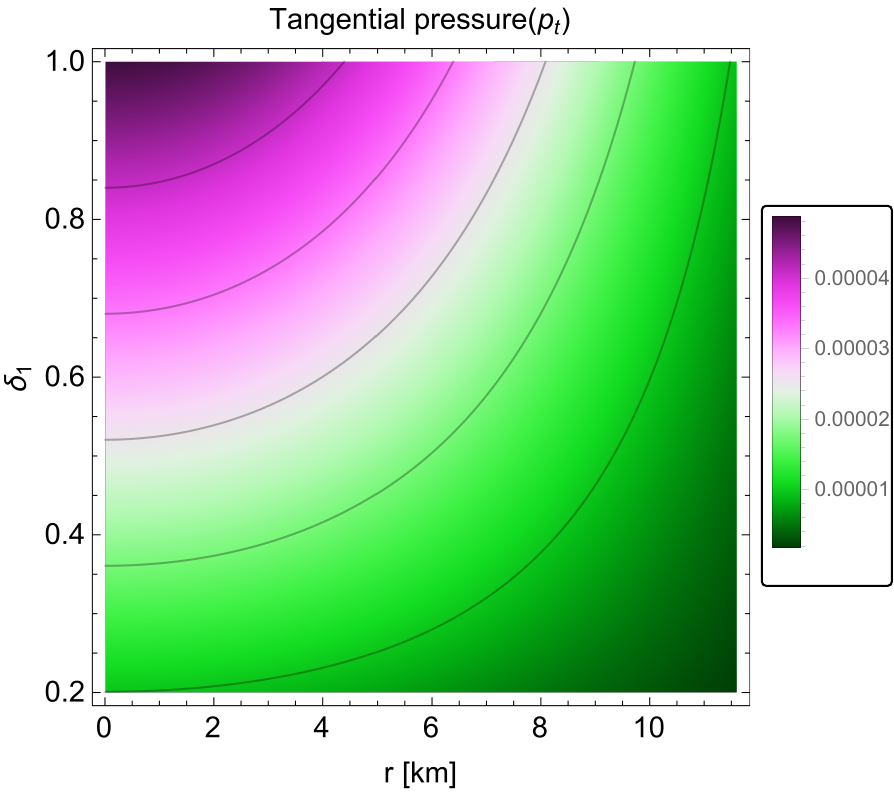}~~~~~\includegraphics[height=7cm,width=7.5cm]{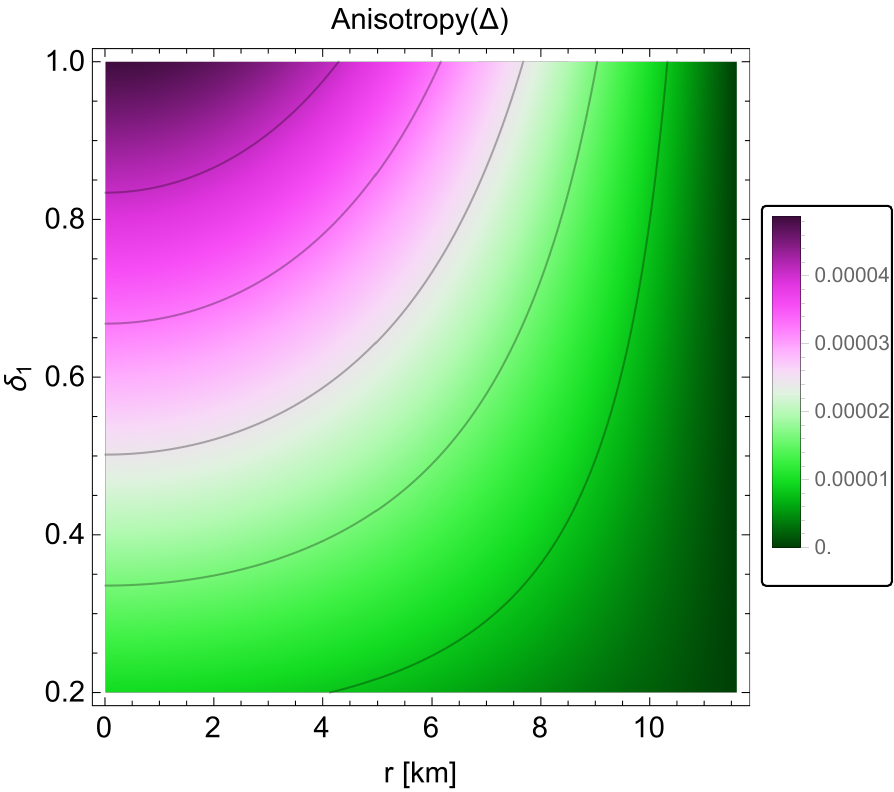}
    \caption{The behaviour of energy density($\rho$), radial pressure($p_{r}$), tangential pressure($p_{t}$) and anisotropy($\Delta$) for a fixed value of $\delta_{2}=0.00001$ versus $\delta_{1}$ and $r$.}
    \label{figure 2}
\end{figure*}

In a static and SS non-ideal fluid system, the anisotropy factor, $\Delta=(p_{t}-p_{r})$, establishes the inequality in pressures, namely  $p_{r}$ and $p_{t}$.
 A force resulting from anisotropic pressure is denoted by $\Delta/r$, which becomes repulsive when $p_{t} > p_{r} \leftrightarrow \Delta > 0$, and attractive when $p_{t} < p_{r} \leftrightarrow \Delta < 0$ in the star model. In the examined matter distribution, when \( p_{t} > p_{r} \), it facilitates the formation of more compact objects in comparison to an isotropic fluid sphere~\cite{Gokhroo:1994fbj,Maurya:2017uyk,maurya2017anisotropic,Maurya:2017jdo,morales2018compact}. The system's stability and equilibrium are enhanced by a positive anisotropy factor. In the center of the star, $\Delta$ has to equal zero. The system remains a repulsive anisotropic force, as shown by Figure~\ref{figure 2}, where the value of the $\Delta$ is positively increasing throughout the star and zero at the center of the star.

\begin{figure*}[!htp]
    \centering
\includegraphics[height=7cm,width=7.5cm]{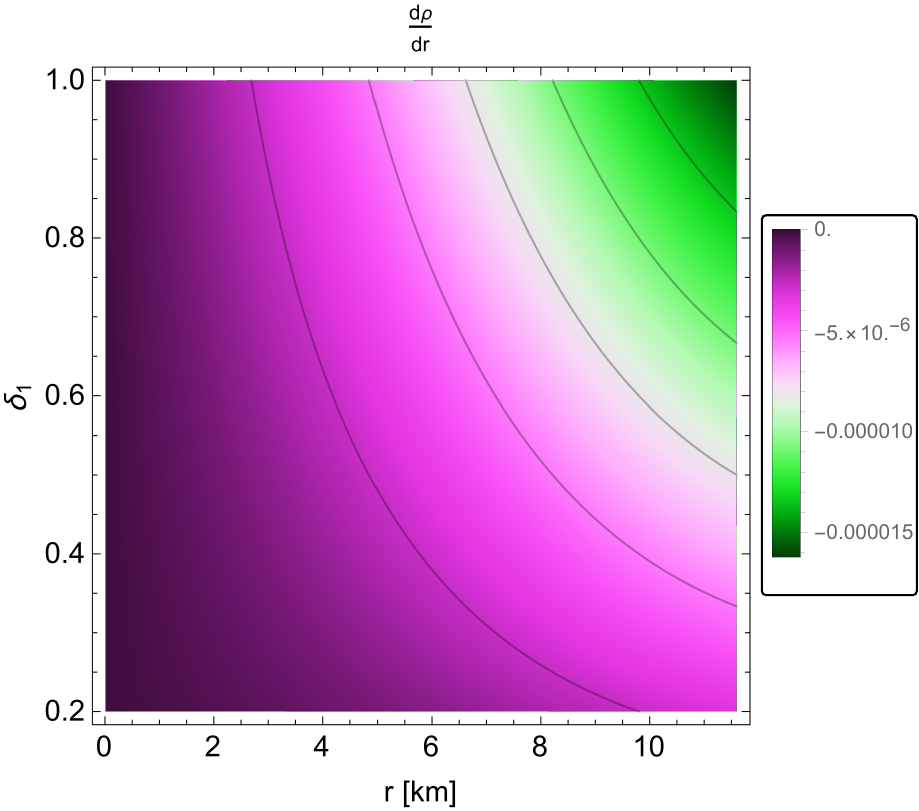}~~~~~~~~~\includegraphics[height=7cm,width=7.5cm]{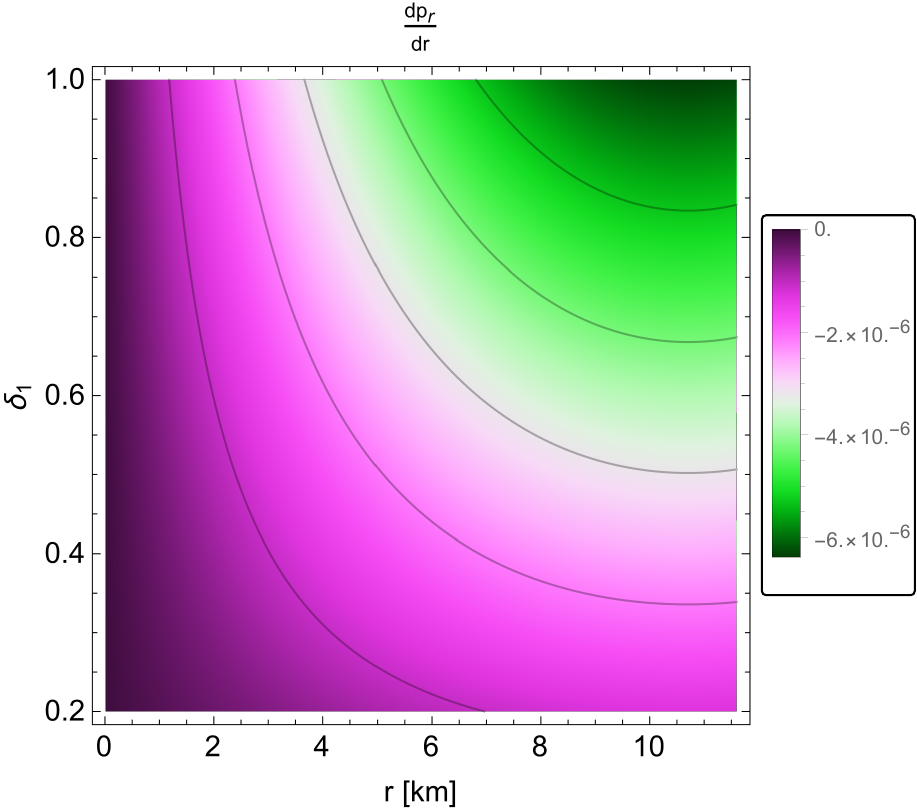}\\~~~~~\includegraphics[height=7cm,width=7.5cm]{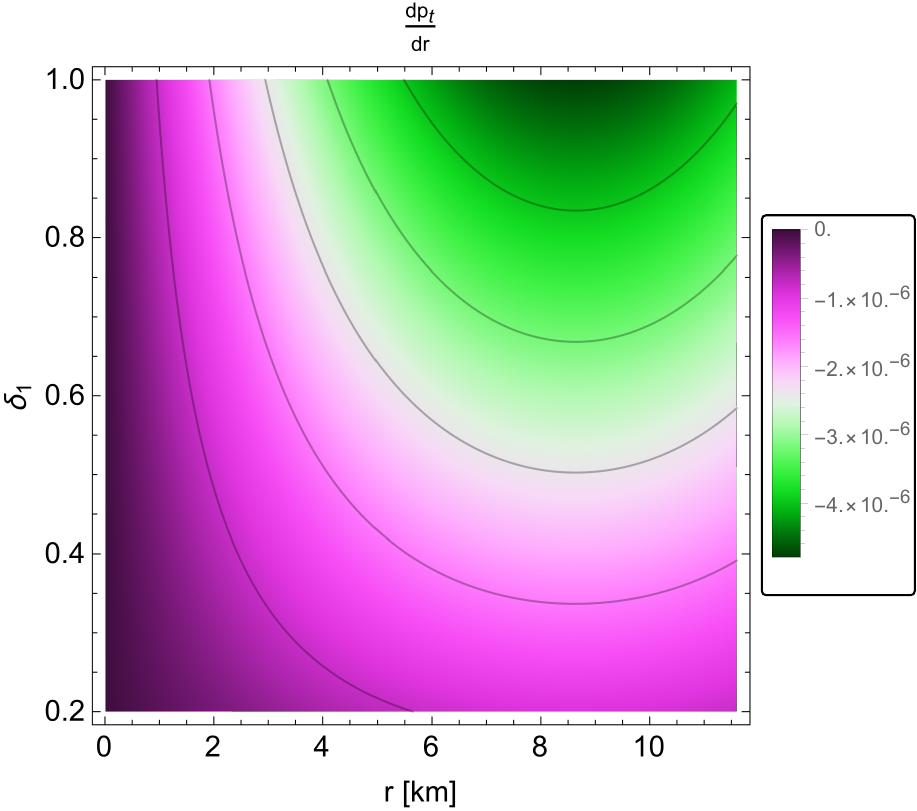}
    \caption{The behaviour of gradient of energy density($\frac{d\rho}{dr}$), gradient of radial pressure($\frac{dp_{r}}{dr}$) and gradient of tangential pressure($\frac{dp_{t}}{dr}$) for a fixed value of $\delta_{2}=0.00001$ versus  $\delta_{1}$ and $r$. }
    \label{figure 3}
\end{figure*}

\subsection{Zel'dovich requirement}\label{sec3.6}
\begin{figure*}[!htp]
    \centering
    \includegraphics[height=7cm,width=7.5cm]{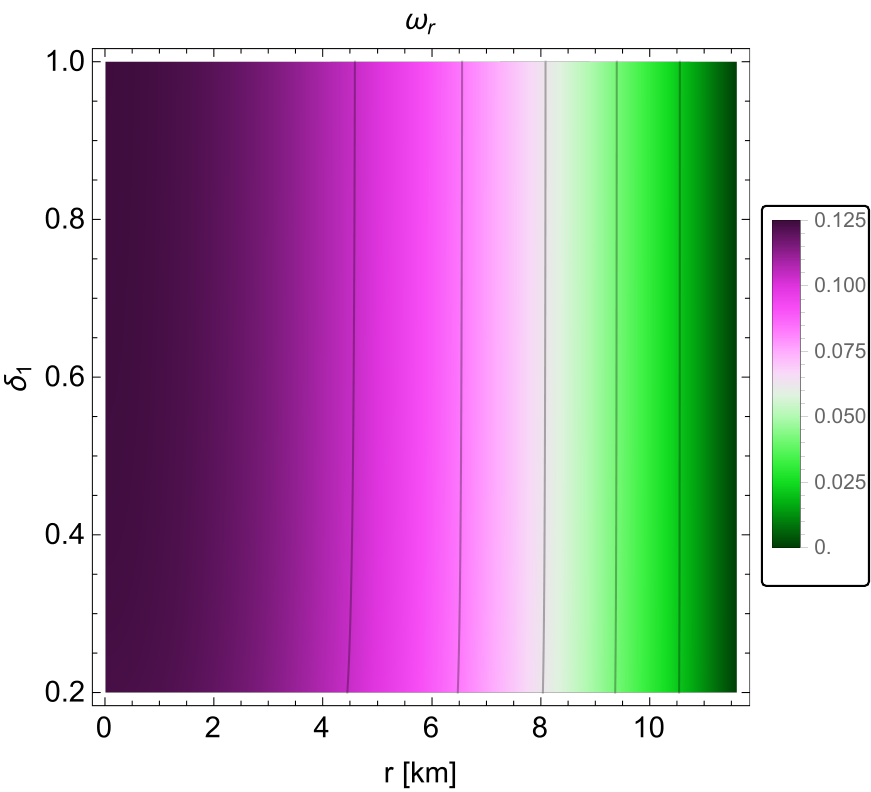}~~~~~\includegraphics[height=7cm,width=7.5cm]{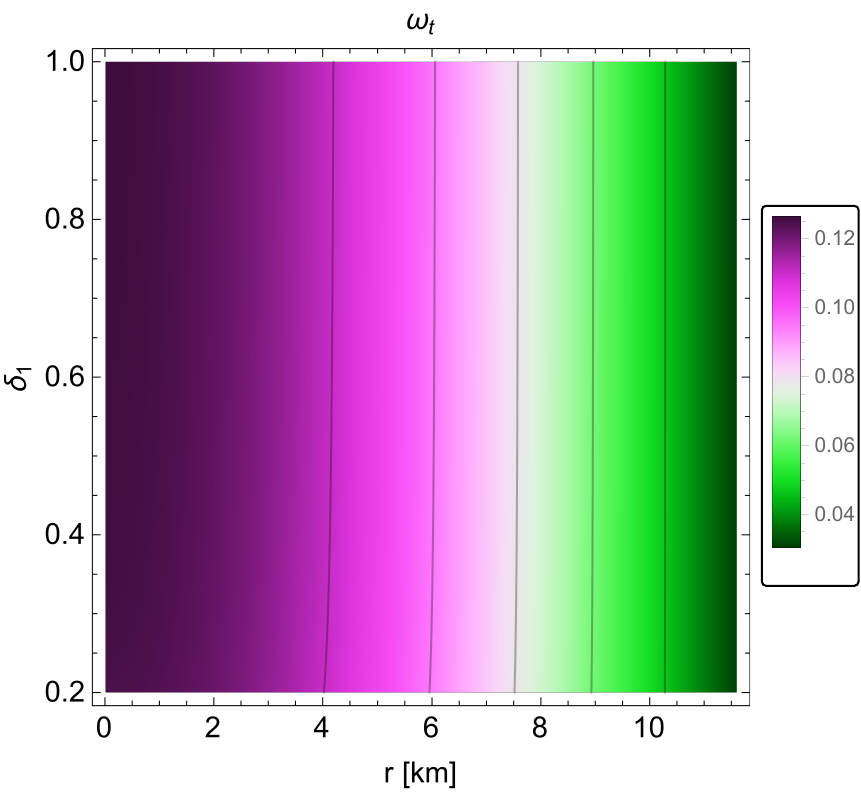} 
    \caption{The behaviour $\omega_{r}$ and $\omega_{t}$ for a fixed value of $\delta_{2} = 0.00001$ versus $\delta_{1}$ and $r$.}
    \label{figure 4}
\end{figure*}

The Zel`dovich ratio is calculated for both radial and tangential orientations ~\cite {Zeldovich:1968ehl}. The model of the compact star evidently fulfils these two requirements, as shown in Fig.~\ref{figure 4}. Regarding the interior of stars, where $r < R$, we utilise $ \omega_{r} = \frac{p_{r}}{\rho}<1$ and $\omega_{t} = \frac{p_{t}}{\rho}<1$, both of which are less than 1. Consequently, the model of a compact star can be sustained against gravitational collapse with support from degeneracy. The expression derived numerically for $\omega_{r}$ and $\omega_{t}$ are:

\begin{eqnarray}\label{eq50}
\omega_{r}=\Bigg(\frac{8 B_1 b \delta _1 f r^2 \log (e) e^{f r^2+h}+\delta _1 \phi _2}{r^2 \phi _3 \left(e^{f r^2+h}-1\right) \left(e^{f r^2+h}+1\right)}+A\Bigg)B,\\
&&\hspace{-7.5cm}\omega_{t}=\Bigg(\frac{\delta _2 \phi _5 \left(b^4 r^4 \text{csch}^4\left(f r^2+h\right)+1\right)+C}{2 \phi _7 \left(e^{f r^2+h}-1\right)^2 \left(e^{f r^2+h}+1\right)^2}\Bigg)B.\label{eq51}
\end{eqnarray}

\subsection{Energy bounds }\label{Sec3.7}
All energy limitations must be satisfied for the model to be viable during data analysis. NEC denotes the null energy condition, WEC signifies the weak energy condition, DEC indicates the dominant energy condition, SEC denotes the strong energy condition, and TEC illustrates the trace energy condition.
\begin{figure*}[!htp]
    \centering
    \includegraphics[height=8cm,width=8cm]{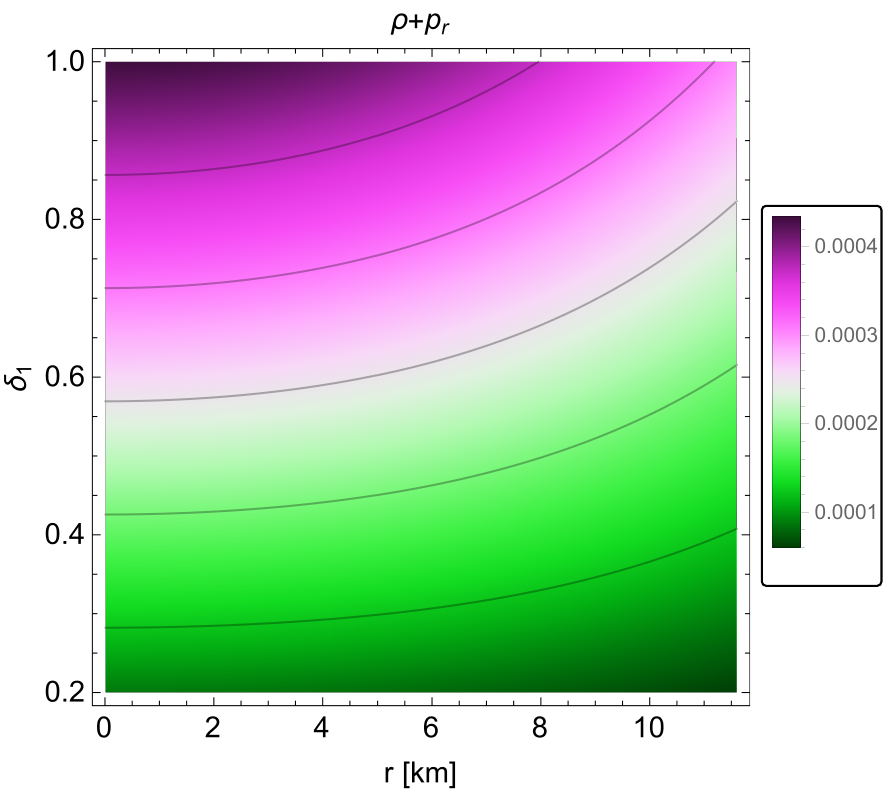}~~~~~\includegraphics[height=8cm,width=8cm]{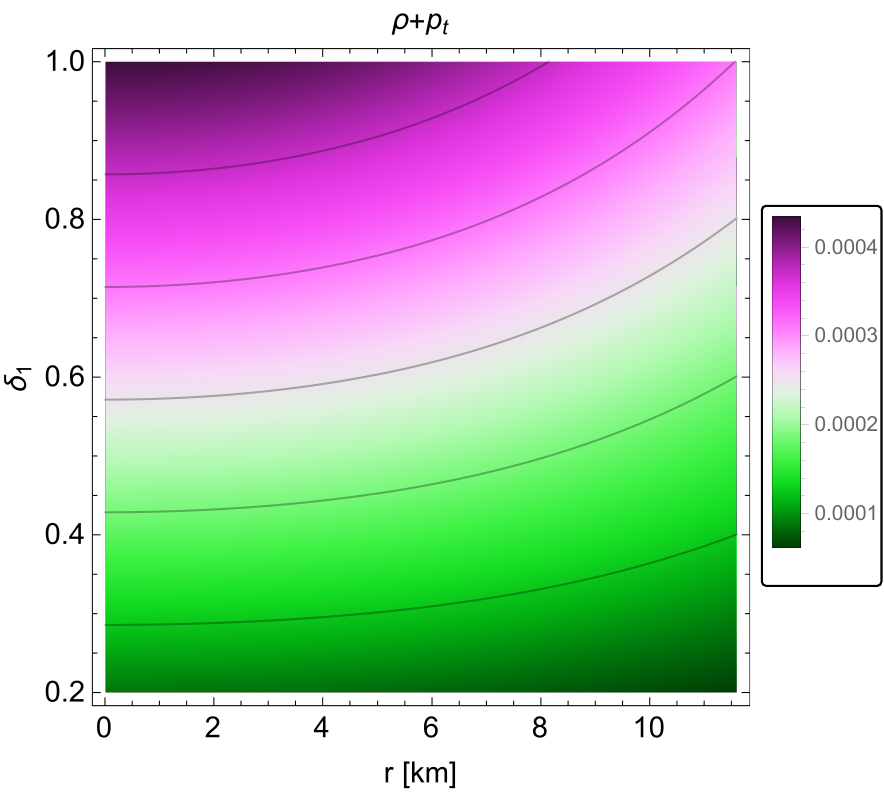}  \\~~~~~\includegraphics[height=8cm,width=8cm]{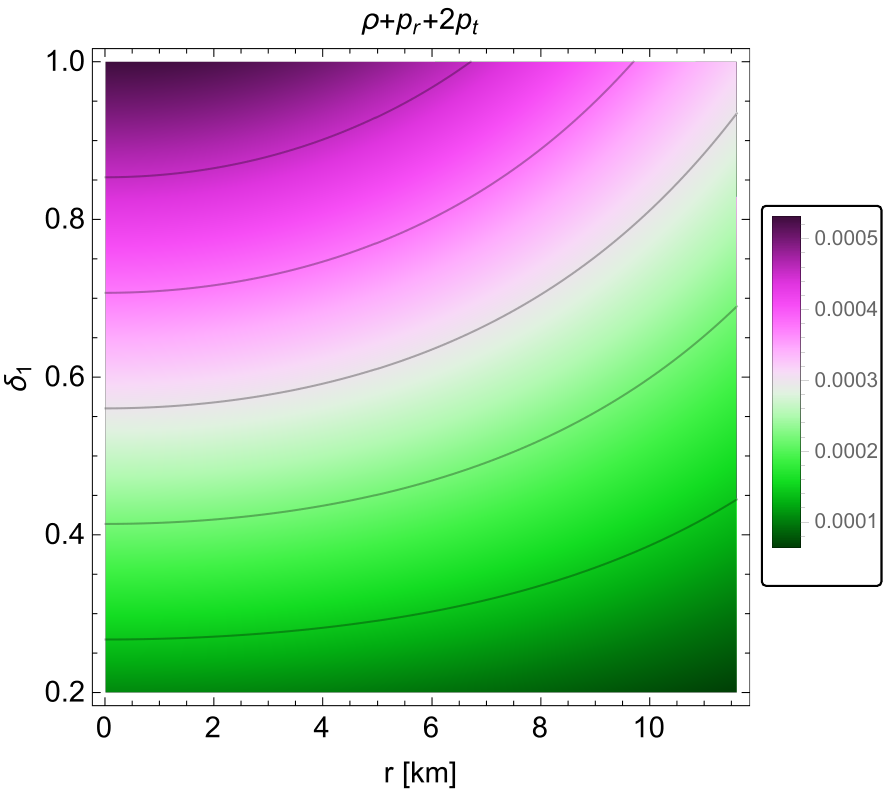}~~~~~\includegraphics[height=8cm,width=8cm]{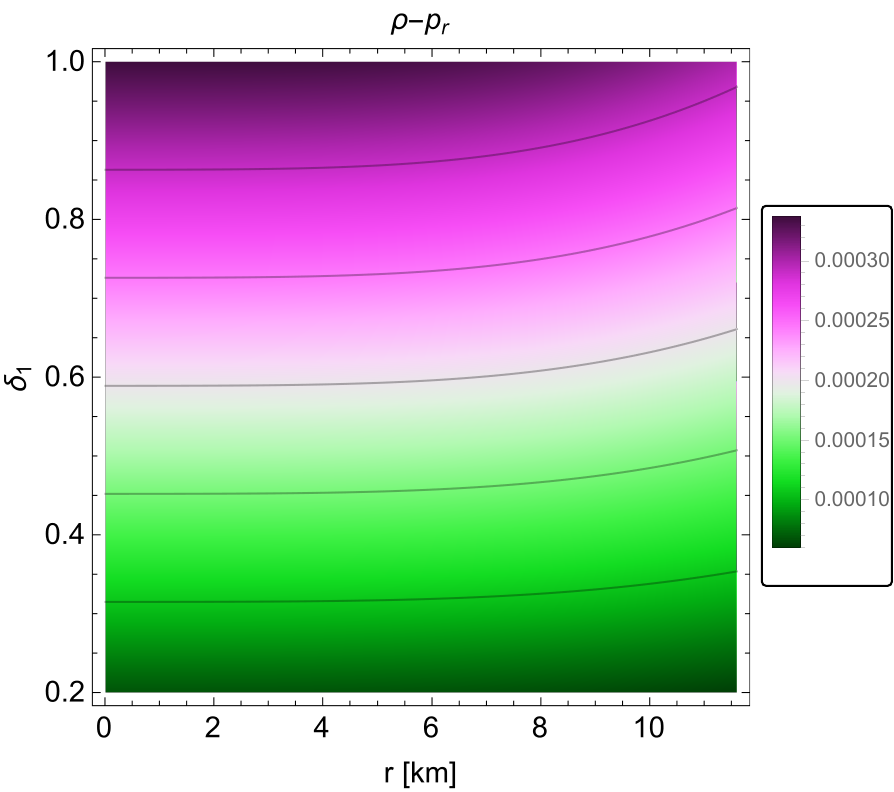}\\~~~~~\includegraphics[height=8cm,width=8cm]{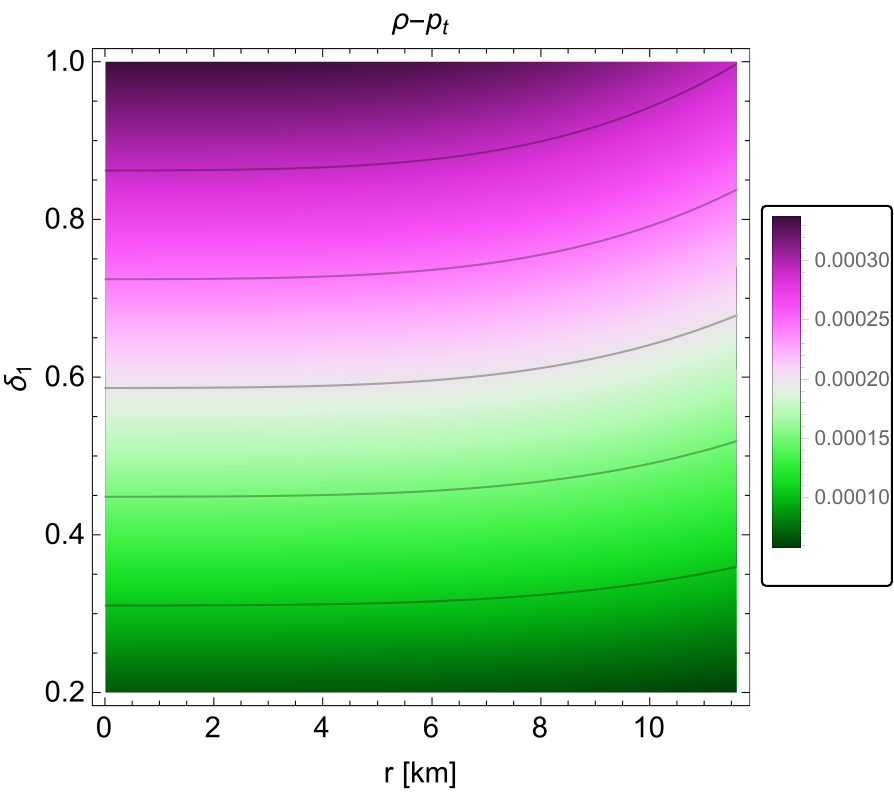}~~~~~\includegraphics[height=8cm,width=8cm]{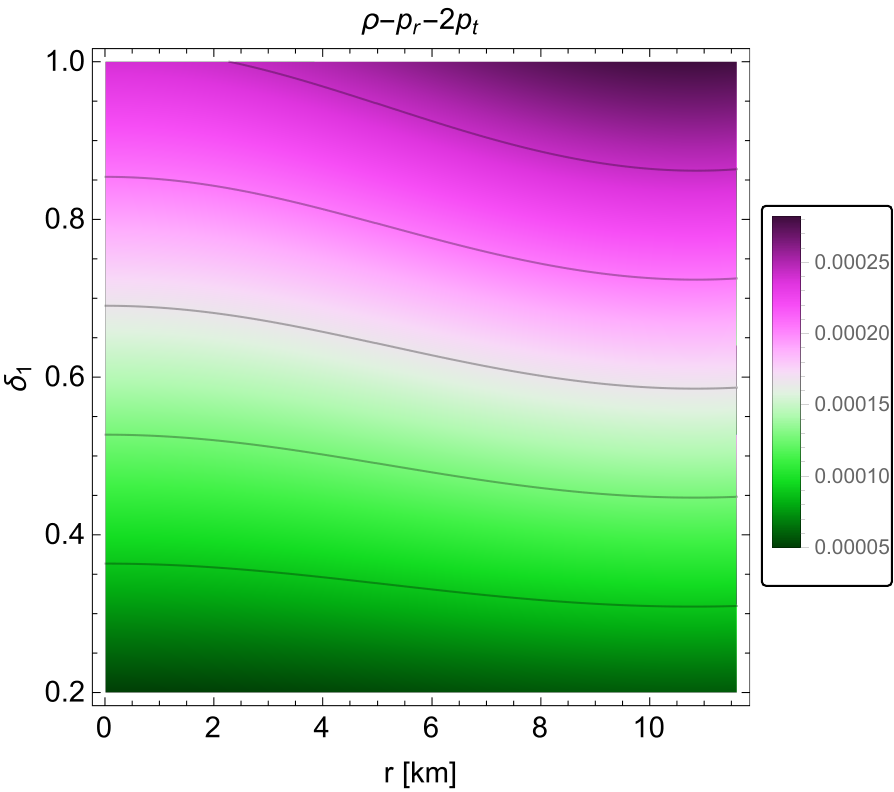} 
    \caption{The behaviour of NEC, WEC, DEC, SEC and TEC for a fixed value of $\delta_{2}=0.00001$  versus $\delta_{1}$ and $r$.}
    \label{figure 5}
\end{figure*}
\begin{eqnarray}\label{eq52}
NEC:\rho \geq 0,\\
WEC:\rho +p_{r}\geq 0,\rho +p_{t}\geq 0,
\label{eq53}\\
DEC:\rho -p_{r}\geq 0,\rho -p_{t}\geq 0,\label{eq54}\\
SEC:\rho+p_{r}+2p_{t} \geq 0,\label{eq55}\\
TEC:\rho-p_{r}-2p_{t}\geq0.\label{eq56}
\end{eqnarray}
Figure~\ref{figure 5} represents the energy requirements graphically; All energy constraints have been determined and are completely satisfied. Consequently, our solutions are physically feasible.\\
The mathematical expression for $\rho$ is given in Eq.~(\ref{eq36}), while the expressions for $\rho + p_{r}$, $\rho + p_{t}$, $\rho - p_{r}$, $\rho - p_{t}$, $\rho + p_{r} + 2p_{t}$, and $\rho - p_{r} - 2p_{t}$ are provided in the appendix.

\subsection{Adiabatic index }\label{Sec3.8}

\begin{figure}[!htp]
    \centering
    \includegraphics[height=7cm,width=7.5cm]{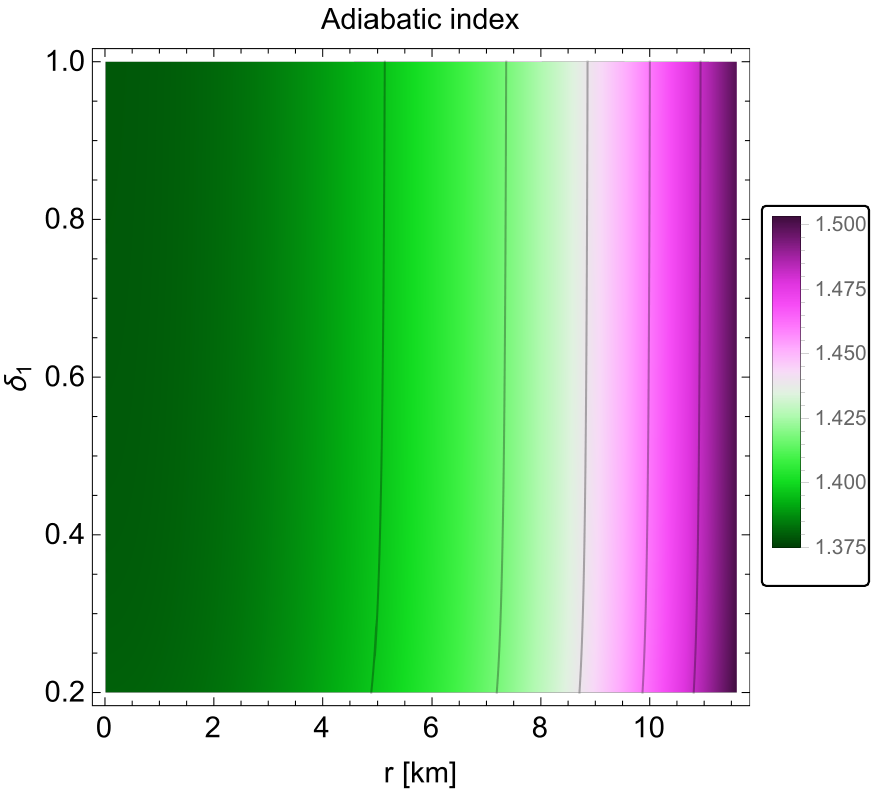}
    \caption{Graphical representation of Adiabatic index $\Gamma$ for a fixed value of $\delta_{2}= 0.00001$ versus $\delta_{1}$ and $r$.}
    \label{figure 6}
\end{figure}

The adiabatic index is an essential parameter that defines the stability of a stellar model. For a SS object, testing the adiabatic index is crucial since it indicates its solidity. For a stable compact star 
 structure the value of the adiabatic index( $\Gamma$) should be greater than $\frac{4}{3}$ to guarantee stability~\cite{heintzmann1975neutron}.The adiabatic index is often defined as:
 \begin{equation}\label{eq57}
\Gamma < \frac{4}{3}-\Bigg[\frac{4}{3}\frac{p_{r}-p_{t}}{\vert p_{r}^{\prime}\vert r}\Bigg]_{max}
 \end{equation}
 Our present investigation examines the graphical behaviour of the adiabatic index, shown in  Fig.~\ref{figure 6}. The value of the adiabatic index($\Gamma$) is greater than $\frac{4}{3}$ for any value of r. From this, we can say that our proposed model satisfies the required stability criterion($\Gamma>\frac{4}{3}$).
 The mathematical expression for the adiabatic index($\Gamma$) is given in the appendix.
 
\subsection{Stability and Equilibrium }\label{Sec3.9}
This subsection discusses the graphical depiction of stability and equilibrium in the stellar model utilising TOV equations, causality Condition, Stability Versus Convection and Harrison-Zeldovich-Novikov Condition in the interior of a compact star.

\subsubsection{Equilibrium requirement}\label{Sec3.9.1}

The Tolman-Oppenheimer-Volkoff (TOV) equation has been described as an equilibrium condition for a compact star, combining gravitational, hydrostatic and anisotropic forces ~\cite{zubair2016possible}. The standard representation of the TOV equation is expressed as
\begin{equation}\label{eq58}
\frac{dp_{r}}{dr}+\frac{\Upsilon^{\prime}(\rho + p_{r})}{2}-\frac{2(p_{t}-p_{r})}{r}=0
\end{equation}

On the other hand, it can be stated as\\\\
$F_{g}+F_{h}+F_{a} = 0$;
$F_{g}=-\frac{\Upsilon^{\prime}(\rho + p_{r})}{2}$,\quad $F_{h}=-\frac{dp_{r}}{dr}$,\quad $F_{a}=\frac{2(p_{t}-p_{r})}{r}.$\\\\

\begin{figure*}[!htp]
    \centering
    \includegraphics[height=8cm,width=11cm]{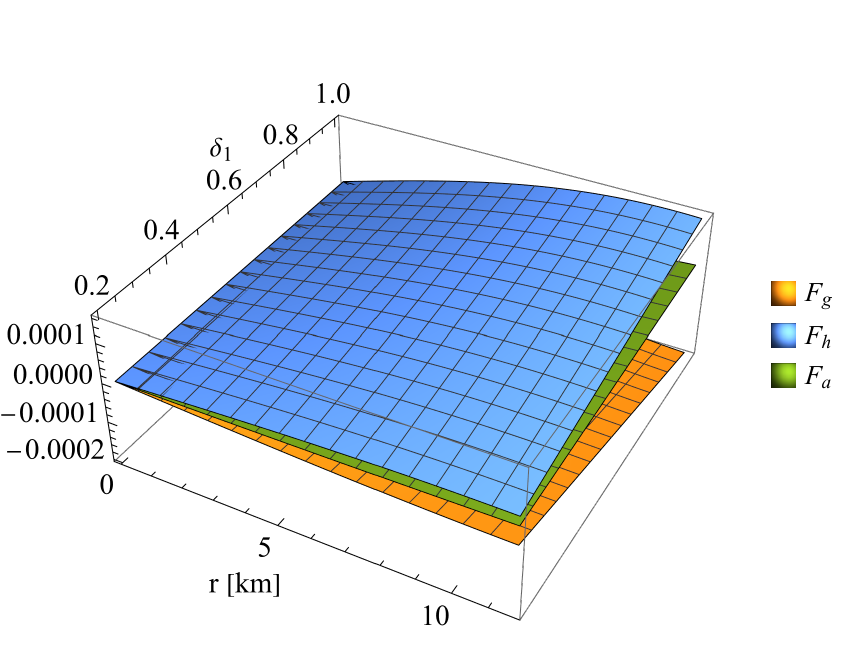}
    \caption{The behaviour of gravitational force ($F_{g}$), hydrostatic force ($ F_{h}$) and anisotropic force ($F_{a}$) versus $r$ and $\delta_{1}$ for a fixed value of $\delta_{2}=0.00001$.}
    \label{figure 7}
\end{figure*}

This $ F_{g}$, $F_{h}$ and $F_{a}$ indicate the gravitational, hydrostatic, and anisotropic forces, respectively. Using the value of  $\rho$, $p_{r}$ and $p_{t}$ in the TOV equation, we represent these forces visually shown in Fig.~\ref{figure 7}. Each of these forces has small positive or negative values; nevertheless, together, they satisfy the TOV equation. Therefore, we find that the revealed stellar structure is in a condition of equilibrium as a result of these forces.\\
The mathematical expression for $ F_{g}$, $F_{h}$ and $F_{a}$ are given in the appendix.

\subsubsection{ Causality and  Stability requirement}\label{Sec3.9.2}
\begin{figure*}[!htp]
    \centering
    \includegraphics[height=7cm,width=7.5cm]{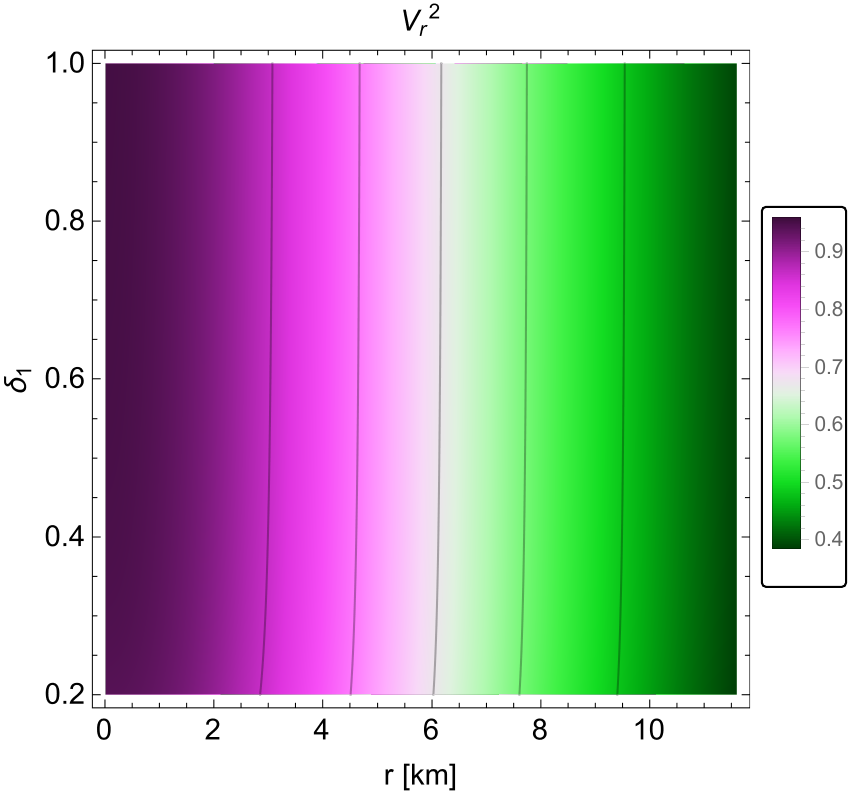}~~~~~\includegraphics[height=7cm,width=7.5cm]{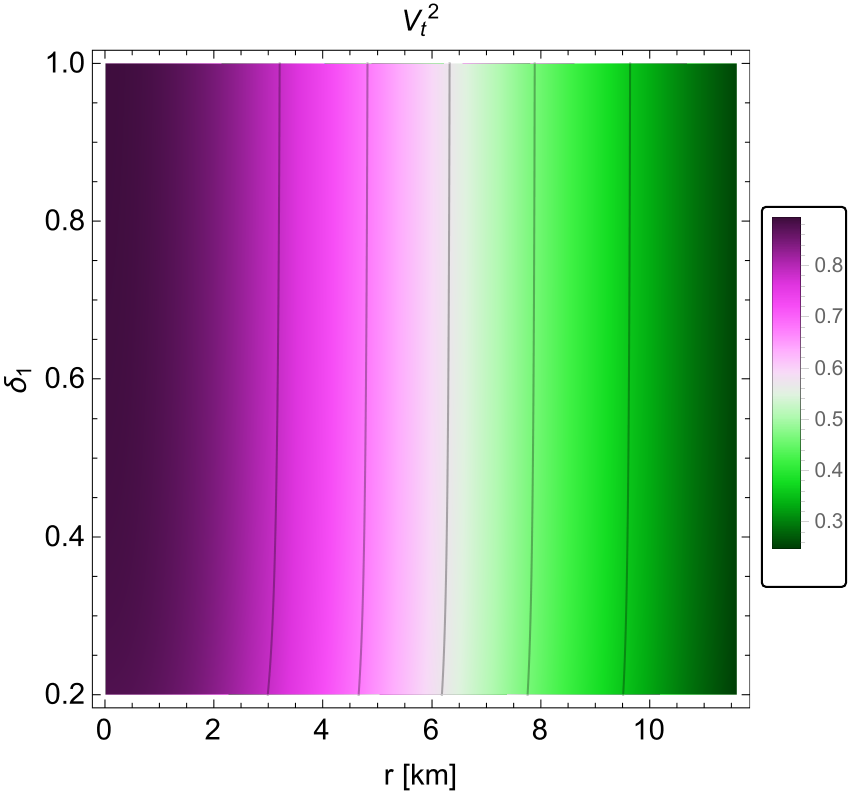}
    ~~~~~\includegraphics[height=7cm,width=7.5cm]{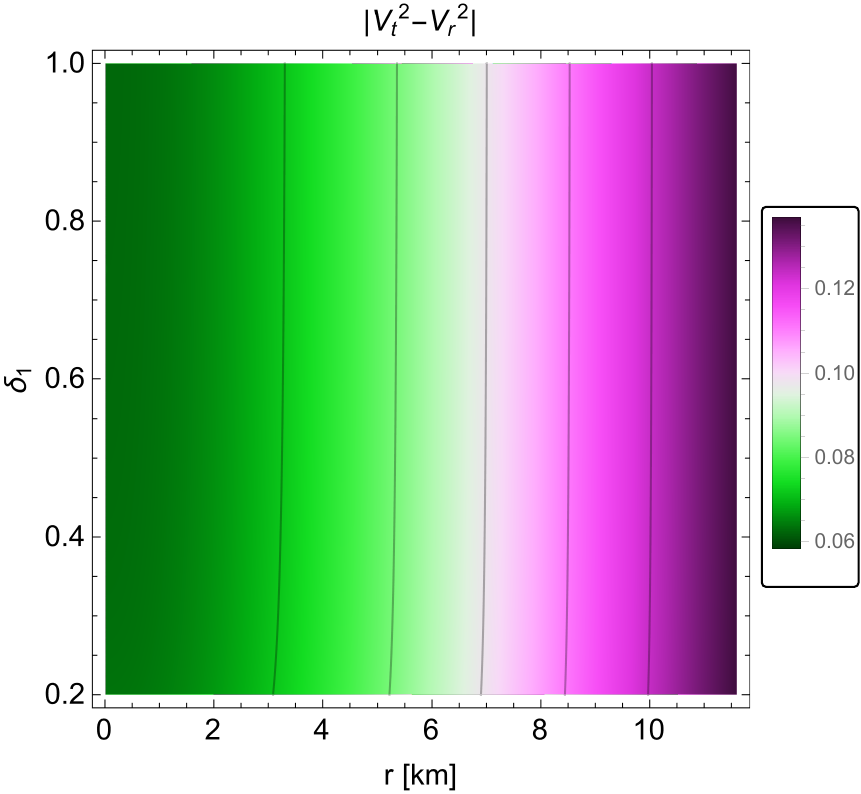} 
    \caption{The behaviour of radial velocity ($V^{2}_{r}$), Tangential velocity ($ V^{2}_{t} $ ) and Resultant velocity($|V^{2}_{t}-V^{2}_{r}|$) for a fixed value of $\delta_{2} = 0.00001$ versus  $\delta_{1}$ and $r$.}
    \label{figure 8}
\end{figure*}

This study examines the validity of the causality criterion for the proposed stellar model. The causality conditions provide boundaries on the radial and tangential velocities, denoted as $V^{2}_{r}$ and $V^{2}_{t}$, respectively. The boundaries are specified as $0<\vert V^{2}_{i} \vert<1$, where $i=r,t$. The $V^{2}_{r}$
and $V^{2}_{t}$ are defined by
\begin{equation}\label{eq59}
V^{2}_{r} = \frac{dp_{r}}{d\rho},\quad V^{2}_{t} = \frac{dp_{t}}{d\rho}
\end{equation}
The graphical representation of these velocities is illustrated in Fig.~\ref{figure 8}, demonstrating the validity of causality criteria.
Furthermore, ~\cite{Abreu:2007ew} has proposed an additional stability requirement for these velocities, denoted
as $-1 \leq V^{2}_{t}-V^{2}_{r} \leq 1 $. This behaviour is regarded as a
significant characteristic of compact star models. In the
present case, we visually examine this state as shown in
Fig.~\ref{figure 8}. The graph strongly illustrates that our model
fulfils this stability criteria as well.\\
The mathematical expression for $V^{2}_{r} = \frac{dp_{r}}{d\rho}$ and $V^{2}_{t} = \frac{dp_{r}}{d\rho}$ are given in the appendix.

\subsubsection{ Stability Versus Convection}\label{Sec3.9.3}

\begin{figure*}[!htp]
    \centering
\includegraphics[height=7cm,width=7.5cm]{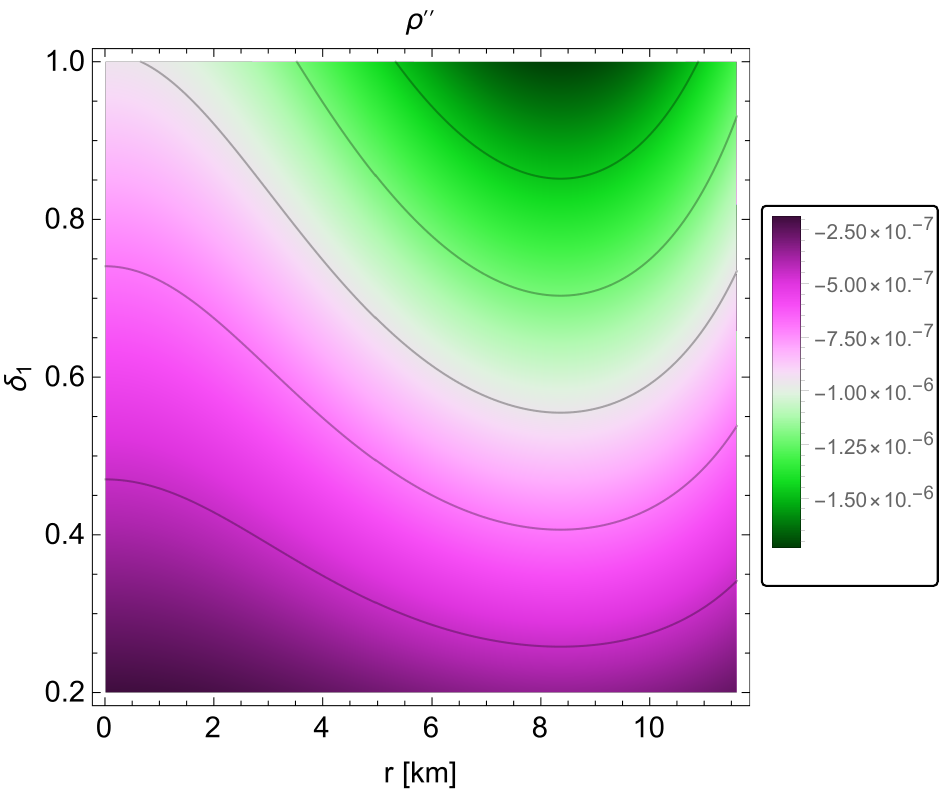}
    \caption{The behaviour of $\rho^{\prime \prime}$ for a fixed value of $\delta_{2}=0.00001$ versus  $\delta_{1}$ and $r$. }
    \label{figure 9}
\end{figure*}

The buoyancy principle inside a fluid demonstrates that every displaced fluid component will revert to its original position, as evidenced by the stability of a self-gravitating sphere in contrast to convection. It was illustrated in ~\cite {Hernandez:2018sjc}, such that
\begin{equation}\label{eq60}
\rho^{\prime \prime}(r) \leq 0
\end{equation}
We demonstrate the graphical representation of $\rho^{\prime \prime}$ as a function of r in Fig.~\ref{figure 9}. The model demonstrates stability in the inner shells during convective motion, whereas the outside shells remain unstable ~\cite {Andrade:2022idg}.\\
The mathematical expression for $\rho^{\prime \prime}$ is given in the appendix.

\subsubsection{ Harrison-Zeldovich-Novikov Stability Condition}\label{Sec3.9.4}
\begin{figure*}[!htp]
    \centering
    \includegraphics[height=7cm,width=7.5cm]{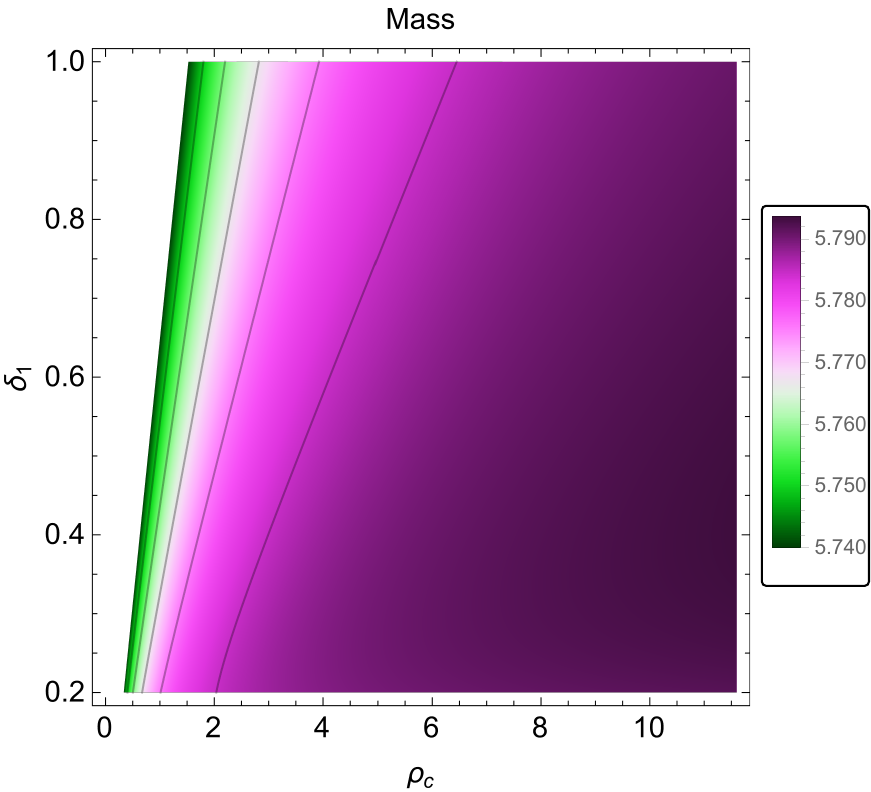}~~~~~\includegraphics[height=7cm,width=7.5cm]{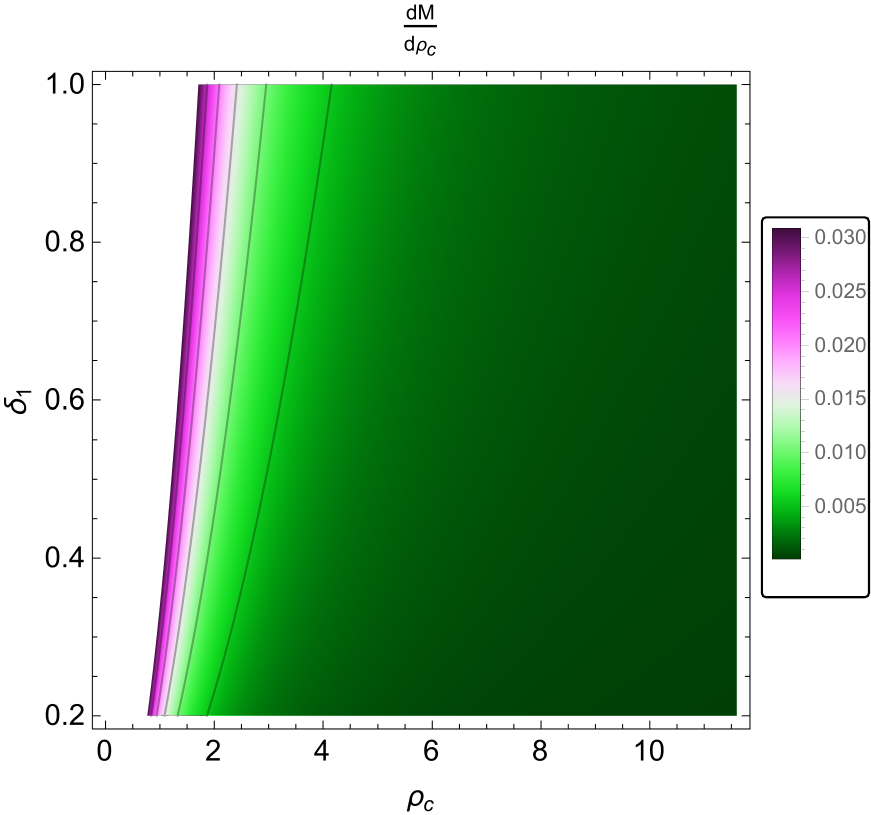}
    \caption{The behaviour of Mass and $\frac{dM}{d\rho_{c}}$ for a fixed value of $\delta_{2}=0.00001$ versus $\delta_{1}$ and the central density $\rho_{c}$ of the proposed compact star.}
    \label{figure 10}
\end{figure*}

Depending on the star's mass and center density, Harrison ~\cite {harrison1965gravitation} and Zeldovich-Novikov ~\cite {zeldovich1971relativistic} suggested stability requirements for the compact star model. Based on this information, they proposed that $\frac{dM}{d\rho_{c}}>0$ for a stable structure, where M and $\rho_{c}$ stand for the compact star's mass and central density, respectively. The mass function of our solution can be defined using the center density as 

\begin{equation}\label{eq61}
M(\rho_{c})=\frac{R}{2}\left(-\frac{D}{E \left(2 \rho _c+\delta _2\right)+D}-\frac{\delta _2 R^2}{6 \delta _1}+1\right).
\end{equation}
When differentiating Eq.~(\ref{eq61}) with respect to $\rho_{c}$, we obtain
\begin{equation}\label{eq62}
\frac{dM}{d\rho_{c}}=\frac{4 E D R}{\left(4 E \rho _c+2 E\delta _2+12 \delta _1 \text{csch}^2(h)\right){}^2}.
\end{equation}
Figure~\ref{figure 10} illustrates that our anisotropic models satisfy the Harrison-Zeldovich-Novikov stability criteria. The stability of our constructions improves with larger radii and stays constant after the central densities reach their peak levels. 

\begin{table*}[!htp]
\centering
\begin{tabular}{ccccc}
      \hline
Compact star&\hspace{0.2cm}$\delta_{1}$&\hspace{0.2cm} density value at r=0($gm/cm^{3}$)&\hspace{0.5cm} density value at r=R($gm/cm^{3}$) &\hspace{0.2cm}pressure value at r=0($dyne/cm^{2}$)\\
    \hline
  Cen X-3&\hspace{0.2cm}0.2&\hspace{0.3cm}0.041349 $\times10^{14}$&\hspace{0.5cm}0.03222$\times10^{14}$&\hspace{0.2cm}0.043506$\times10^{34}$\\
    \hline
    Cen X-3&\hspace{0.2cm}0.4&\hspace{0.2cm}0.082698$\times10^{14}$&\hspace{0.5cm}0.06444$\times10^{14}$ &\hspace{0.2cm}0.091846$\times10^{34}$\\
    \hline
   Cen X-3&\hspace{0.2cm}0.6&\hspace{0.2cm}0.124047$\times10^{14}$ &\hspace{0.6cm}0.09666 $\times10^{14}$&\hspace{0.3cm}0.140186 $\times10^{34}$\\
    \hline 
    Cen X-3&\hspace{0.2cm}0.8&\hspace{0.2cm}0.165396$\times10^{14}$ &\hspace{0.5cm}0.12888$\times10^{14}$ &\hspace{0.2cm}0.188526$\times10^{34}$ \\
    \hline
   Cen X-3&\hspace{0.2cm}1.0 &\hspace{0.2cm}0.206745$\times10^{14}$&\hspace{0.5cm}0.16110$\times10^{14}$&\hspace{0.2cm}0.236866$\times10^{34}$ \\
    \hline
    \end{tabular}
    \caption{The value of central density($\rho_{c}$), surface density and the central pressure($p_{rc}$) for different values of $\delta_{1}$}
    \label{Tab3}
\end{table*}

\subsection{ Redshift Profile}\label{Sec3.10}
\begin{figure*}[!htp]
    \centering
    \includegraphics[height=7cm,width=7.5cm]{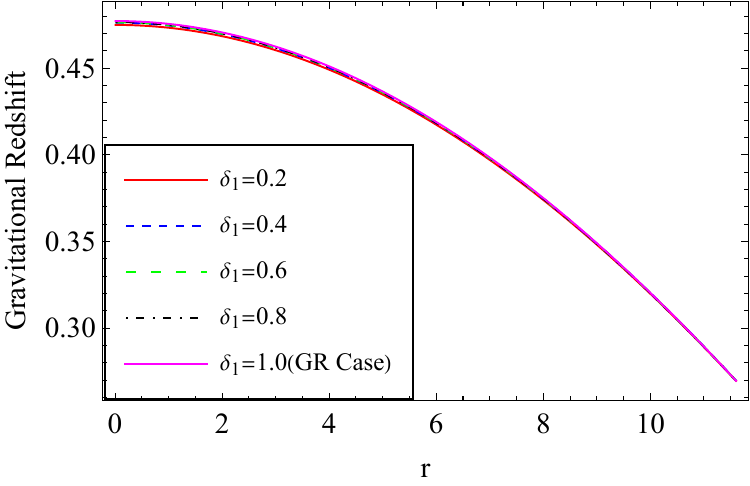}~~~~~\includegraphics[height=7cm,width=7.5cm]{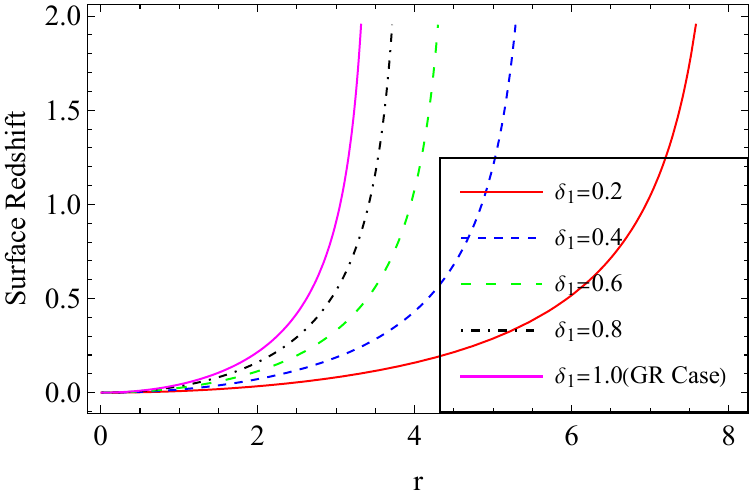}
    \caption{The behaviour of Gravitational Redshift ($Z_{G}$) and Surface Redshift ($Z_{S}$) for a fixed value of $\delta_{2}=0.00001$ versus $\delta_{1}$ and $r$.}
    \label{figure 11}
\end{figure*}

Gravitational redshift is a phenomena in which electromagnetic waves or photons lose energy as they ascend from a gravitational well. Electromagnetic radiation grows its wavelength and diminishes photon energy when it exits a gravitational well. Since it must propagate simultaneously at the speed of light, it consumes energy by changing its frequency rather than its velocity. When energy is reduced, the photon's wave frequency decreases and its wavelength extends, resulting in a shift toward the red end of the electromagnetic spectrum, known as redshift. The behaviour of interior redshift similarly reflects the profile of interior density. A photon must traverse a longer distance and pass through a denser part of the compact star's interior to exit the centre and reach the surface. This greater dispersion causes a significant energy loss. A photon that ascends near the surface will traverse a shorter distance through a less dense region, leading to less dispersion and energy loss. The inner redshift is maximal at the center and minimal at the surface. The surface redshift is contingent upon the surface gravity, which is determined by the radius and total mass of the star object. Surface gravity rises due to a little augmentation in radius, equivalent to the growth in mass. Consequently, the surface redshift escalates monotonically approaching the star's surface. Gravitational and surface redshifts are characterized as follows:
\begin{equation}\label{63}
 Z_{G}= \frac{1}{\sqrt{\vert e^{\Upsilon(r)}\vert}}-1 = \frac{1}{A_{1}+\frac{B_{1}b}{2 f}\log \Bigg(\frac{e^{(fr^{2}+h)}-1}{e^{(f r^{2}+h)}+1}\Bigg)}-1,
 \end{equation}\\
 \begin{equation}\label{64}
Z_{S}= \frac{1}{\sqrt{1-2u(r)}} -1.
\end{equation}
 The value of $u(r)$ is specified in Eq.~(\ref{eq68}), indicating that the gravitational redshift monotonically decreases approaching the surface, whereas surface redshift monotonically increases when we move to surface, as seen in Fig.~\ref{figure 11}. 
 
\subsection{Mass and compactness of the model}\label{Sec3.11}
\begin{figure*}[!htp]
    \centering
\includegraphics[height=7cm,width=7.5cm]{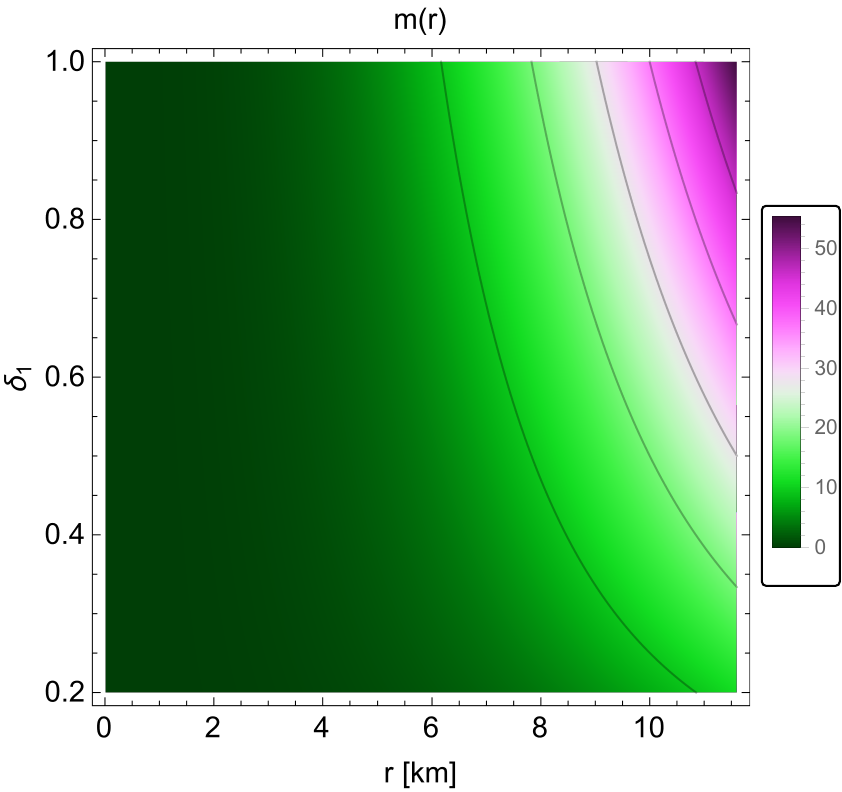}~~~~~\includegraphics[height=7cm,width=7.5cm]{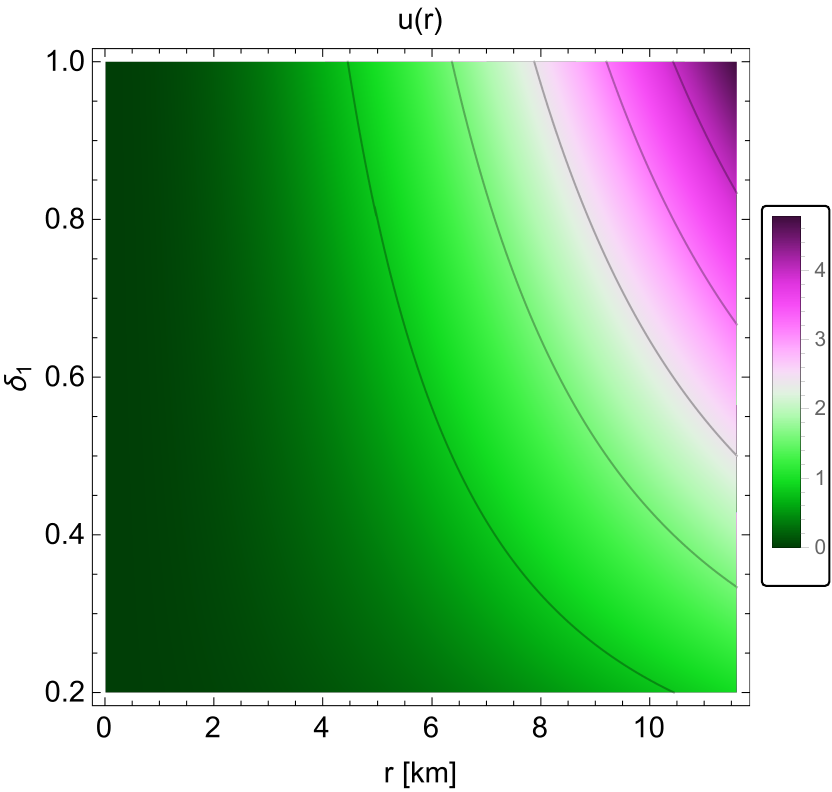}
    \caption{The behaviour of $m(r)$ and $u(r)$ versus and $r$ and  $\delta_{1}$ for a fixed value of $\delta_{2}=0.00001$.}
    \label{figure 12}
\end{figure*}

The mass and compactness of the compact star are shown in Fig.~\ref{figure 12}. It is possible to see how parameter $\delta_{1}$ affects the compactness and mass of the compact star. Increasing $\delta_{1}$ results in an increase in mass and compactness. The compact star's mass and compactness values are zero in its core and increase monotonically as it moves toward the surface. It's important to remember that every star structure has a mass-radius ratio constraint, which Buchdahl provides and is $\frac{2M}{R} \leqslant \frac{8}{9}$. In Ref.~\cite{Buchdahl:1959zz}, this limit is shown in Table~\ref{Table1} and has been calculated for the our compact star model. This constraint is inferior than $\frac{4}{9}$, indicating that our model adheres to Buchdahl's criteria. The functions $m(r)$ and $u(r)$, corresponding to the radial coordinate $r$, reflect the mass and compactness of the compact star, respectively, as shown below: 

\begin{eqnarray}\label{eq65}
&& \hspace{-0.5cm}m(r)=\frac{\kappa}{2}\int^{r}_{0}\rho r^{2}dr
,\\ 
&& \hspace{-0.5cm}
u(r)=\frac{\kappa}{2r}\int^{r}_{0}\rho  r^{2}dr.\label{eq66}
\end{eqnarray}
The following is the expression for the mass and compactness of the compact star:
\begin{eqnarray}\label{eq67}
    && \hspace{-0.5cm}m(r)=-\frac{2 \pi  r^3 \left(2 b^2 \left(\delta _2 r^2-6 \delta _1\right)+\delta _2 \left(\phi _{14}-1\right)\right)}{3 \left(2 b^2 r^2+\phi _{14}-1\right)},\\
&& \hspace{-0.5cm}
u(r) = -\frac{2 \pi  r^2 \left(2 b^2 \left(\delta _2 r^2-6 \delta _1\right)+\delta _2 \left(\phi _{14}-1\right)\right)}{3 \left(2 b^2 r^2+\phi _{14}-1\right)}\label{eq68}.
\end{eqnarray}
The star acquired here lies inside the threshold of being a compact star, with a solar mass of 1.49 for a value of $\delta_{1}$=1. The graph's upward trend indicates that the star's mass increases progressively with the rising of the parameter $\delta_{1}$. In this context, the constant $\delta_{2}$ is set to 0.00001, thereby exerting little influence on the mass and compactness of the model. 

\section{Conclusion}\label{sec4}
This study examines a SS compact star model with non-perfect fluid. We examined the generic function $\mathcal{F}(\mathcal{Q})$ defined as $\mathcal{F}(\mathcal{Q})=\delta_{1}\mathcal{Q} + \delta_{2} $, where $\delta_{1}$ and $\delta_{2}$ are integration constant. The Karmarkar criteria have been used to develop the differential equation connecting the metric elements of spacetime. By solving this differential equation, we have derived an expression for the metric element $e^{\Upsilon}$ in relation to $e^{\sigma}$. The juxtaposition of the internal spherical metric with the exterior Schwarzschild metric imposed constraints on the arbitrary constants throughout this structure. The results acquired were illustrated visually in Fig.~\ref{figure 1}--~\ref{figure 12}. The main results may be described as follows:\\
\begin{itemize}

\item Figure~\ref{figure 1} illustrates that the graph of $e^{-\sigma}$ is regular, finite, and decreasing towards the surface, whereas the graph of $e^{\Upsilon}$ is regular, finite, and growing towards the surface, with both graphs coinciding at the boundary.\\

\item The energy density and both radial and tangential pressure of the star are positive, finite, and monotonically decreasing throughout its entirety, as seen in Fig.~\ref{figure 2}. The radial pressure diminishes toward the star's boundary, with the central pressure approximating $10^{34} dyne/cm^{2}$ and the central density about $10^{14} g/cm^{3}$, as seen in Table~\ref{Tab3}.\\

\item We observe from Fig.~\ref{figure 2} that the anisotropy factor $\Delta$ vanishes at the centre of the star and remains positive throughout the star, showing the presence of opposing forces. The graphical behaviour of the gradients of density, radial pressure, and tangential pressure is illustrated in Fig.~\ref{figure 3}.

\item  The value of $\omega_{r}$ and $\omega_{t}$ are greater than zero and less than one that is $0<\omega_{r},\omega_{t}<1$ and both the graph are monotonically decreasing towards the surface as shown in Fig.~\ref{figure 4}.

\item The energy criteria, including the null energy condition (NEC), weak energy condition (WEC), dominant energy condition (DEC), strong energy condition (SEC) and trace energy condition (TEC) are fulfilled for our anisotropic models are shown in Fig.~\ref{figure 5}.

\item The adiabatic index $\Gamma$ for our model exceeds 4/3 at all interior points within the star, shown in Fig.~\ref{figure 6} , therefore it verify the stability of our model.

\item We analyze the TOV equation to characterise the equilibrium condition influenced by gravitational force $(F_{g})$, hydrostatic force $(F_{h})$ and anisotropic forces $(F_{a})$. Figure~\ref{figure 7} illustrates that the $(F_{g})$ is counterbalanced by the combined effects of $(F_{h})$ and $(F_{a})$ to achieve the necessary stability of the model. Nevertheless, the impact of $(F_{a})$ is less than that of $(F_{h})$.

\item The radial and tangential sound velocities in our model obey the causality criterion, namely $ V^{2}_{r}, V^{2}_{t} < 1.$ They also obey the Abreu constraint: $-1 \leq V^{2}_{t}-V^{2}_{r} \leq 1 $. The graph of  $ V^{2}_{r}, V^{2}_{t}$ are monotonically decreasing towards the surface as shown in Fig.~\ref{figure 8}. For a stable compact star, the condition for stability against convection requires that the stability criterion be negative throughout the stellar interior. In our model, the value of the stability criterion is negative throughout the star, as shown in Fig.~\ref{figure 9}, indicating that the model is stable against convective instability.\\

\item We observe from Fig.~\ref{figure 10} our model satisfy Harrison-Zeldovich-Novikov condition that is $\frac{dM}{d\rho_{c}}>0$.

\item Figure~\ref{figure 11} illustrates that gravitational redshift exhibits a monotonically declining trend, but surface redshift demonstrates a monotonically growing trend as one approaches the surface. Gravitational redshift is greatest in the centre and smallest at the boundary of the compact star, whereas surface redshift is smallest at the centre and greatest at the boundary of the model. ~\cite{Boehmer:2006ye} assert that the redshift must fulfil the requirement $Z \leq 5$ and in the model of an anisotropic star, Ivanov ~\cite{Ivanov:2002xf} proposed that the surface redshift is limited to a maximum of 3.842 when the pressure \( p_t \) adheres to the SEC and 5.211 when \( p_t \) complies with the DEC. Figure~\ref{figure 11} demonstrates that the Gravitational and Surface redshift of our models are in substantial concordance.

\item However, Fig.~\ref{figure 12} displays the mass and compactness of the compact star. The influence of factor $\delta_{1}$ on the compactness and mass of the star is evident. As the magnitude of $\delta_{1}$ grows, both mass and compactness correspondingly rise. The mass and compactness values are 0 in the center of star and rise gradually approaching its outermost layer. It is important to acknowledge that a mass-radius ratio constraint exists for every star configuration, as established by Buchdahl, which is expressed as $\frac{2M}{R} \leqslant \frac{8}{9}$~\cite{Buchdahl:1959zz}. The bound has been calculated for our star models and is shown in Table~\ref{Table1}. This constraint is inferior than $\frac{4}{9}$, indicating that our model adheres to Buchdahl's criteria. 

\item We have examined the behaviour of various quantities, including metric elements, energy density, radial and tangential pressures, energy and pressure gradients concerning the radial coordinate, anisotropy parameter, $ \omega_{r}$ and $\omega_{t}$,  energy conditions, adiabatic index, TOV equation, causality conditions, Harrison-Zeldovich-Novikov condition, redshift, 
 mass and compactness for the astronomical object Cen X-3. The results are summarized in Table~\ref{Table2}, which confirms the physical plausibility and mathematical coherence of our SS model.

\end{itemize}

\section*{Acknowledgments}
Sat Paul acknowledges the Department of Mathematics, Central University of Haryana, India, for providing the necessary support for writing and finalising the paper. I also acknowledge the Ministry of Tribal Affairs, New Delhi, India, for providing financial support. Jitendra Kumar acknowledges the authority of the Central University of Haryana (CUH). S. K. Maurya thanks the administration of the University of Nizwa in the Sultanate of Oman for their constant support and encouragement.

\section*{Data access statement}
This manuscript includes no accompanied data, therefore the data will not be deposited. The essential calculations and graphic description are already included in the paper.
\section*{Conflict of interest}
In this research paper, The authors have no conflicts of interest.

\begin{widetext}
\section*{Appendix}
\begin{eqnarray*}
&&\hspace{-2.7cm}\phi _1=2 b^2 \text{csch}^2\left(f r^2+h\right) \left(-3 \delta _1+4 \delta _1 f r^2 \coth \left(f r^2+h\right)+\delta _2 r^2\right)\\
&&\hspace{-2.7cm}\phi _2=\left(e^{2 \left(f r^2+h\right)}-1\right) \left(2 A_1 f+B_1 b \log \left(\frac{e^{f r^2+h}-1}{e^{f r^2+h}+1}\right)\right)\\
&&\hspace{-2.7cm}\phi _3=\left(b^2 r^2 \text{csch}^2\left(f r^2+h\right)+1\right) \left(2 A_1 f+B_1 b \log \left(\frac{e^{f r^2+h}-1}{e^{f r^2+h}+1}\right)\right)\\
&&\hspace{-2.7cm}\phi _4=16 B_1 b \delta _1 f \log (e) e^{f r^2+h} \left(e^{2 \left(f r^2+h\right)}-1\right)-16 B_1 b \delta _1 f^2 r^2 \log ^2(e) e^{f r^2+h} \left(e^{2 \left(f r^2+h\right)}+1\right)\\
&&\hspace{-2.7cm}\phi _5=\left(e^{ 2\left(f r^2+h\right)}-1\right)^2 \left(2 A_1 f+B_1 b \log \left(\frac{e^{f r^2+h}-1}{e^{f r^2+h}+1}\right)\right)\\
&&\hspace{-2.7cm}\phi _6=\text{csch}^2\left(f r^2+h\right) \left(4B_1 b \delta _1 f r^2 \log (e) \left(e^{2 \left(f r^2+h\right)}-1\right) e^{f r^2+h}-8 B_1 \phi _8+2 \delta _1 \phi _9-\phi _5 \left(\delta _1-\delta _2 r^2\right)\right)\\
&&\hspace{-2.7cm}\phi _7=\left(b^2 r^2 \text{csch}^2\left(f r^2+h\right)+1\right)^2 \left(2 A_1 f+B_1 b \log \left(\frac{e^{f r^2+h}-1}{e^{f r^2+h}+1}\right)\right)\\
&&\hspace{-2.7cm}\phi _8=b \delta _1 f^2 r^4 \log ^2(e) e^{f r^2+h} \left(e^{2 \left(f r^2+h\right)}+1\right)\\
&&\hspace{-2.7cm}\phi _9=2 \delta _1 f r^2 \left(e^{2 \left(f r^2+h\right)}-1\right) \coth \left(f r^2+h\right) \left(4 B_1 b f r^2 \log (e) e^{f r^2+h}+\phi _2\right)\\
&&\hspace{-2.7cm}\phi _{10}=\left(e^{2 \left(f R^2+h\right)}-1\right) \log \left(\frac{e^{f R^2+h}-1}{e^{f R^2+h}+1}\right)\\
&&\hspace{-2.7cm}\phi _{11}=b^2 \left(2 \delta _1-\delta _2 R^2\right) \text{csch}^2\left(f R^2+h\right)\\
&&\hspace{-2.7cm}\phi _{12}=\delta _1-b^2 \left(2 \delta _1-\delta _2 R^2\right) \text{csch}^2\left(f R^2+h\right)\\
&&\hspace{-2.7cm}\phi _{13}=\sqrt{\frac{1}{b^2 R^2 \text{csch}^2\left(f R^2+h\right)+1}} \left(\delta _2-b^2 \left(2 \delta _1-\delta _2 R^2\right) \text{csch}^2\left(f R^2+h\right)\right)\\
&&\hspace{-2.7cm}\phi _{14}=\cosh \left(2 \left(f r^2+h\right)\right)\\
&&\hspace{-2.7cm}A=\frac{\delta _2}{2}-\frac{\delta _1}{r^2}\\
&&\hspace{-2.7cm}B=-\frac{2 \left(b^2 r^2 \text{csch}^2\left(f r^2+h\right)+1\right)^2}{b^4 r^2 \left(\delta _2 r^2-2 \delta _1\right) \text{csch}^4\left(f r^2+h\right)+\delta _2+\phi _1}\\
&&\hspace{-2.7cm}C=2 b^2 \phi _6+\phi _4\\
&&\hspace{-2.7cm}D=6 \delta _1 \text{csch}^2(h)\\
&&\hspace{-2.7cm}E=R^2 \text{csch}\left(f R^2+h\right) \text{csch}\left(f R^2+h\right)\\
&&\hspace{-2.7cm} \phi_{15}= 1+b^{2}r^{2}\text{csch}\left(f r^2+h\right)^{2}\\
&&\hspace{-2.7cm}\phi_{16}= a B_1 \delta _1 \left(\frac{2 f r \log (e) e^{f r^2+h}}{e^{f r^2+h}+1}-\frac{2 f r \log (e) e^{f r^2+h} \left(e^{f r^2+h}-1\right)}{\left(e^{f r^2+h}+1\right)^2}\right) \left(8 b B_1 f r^2 \log (e) e^{f r^2+h}+\phi _2\right)\\
&&\hspace{-2.7cm} \phi_{18}=
\frac{2 \delta _1 f \log (e) e^{f r^2+h} \left(8 b B_1 f r^2 \log (e) e^{f r^2+h}+\phi _2\right)}{r \phi _3 \left(e^{f r^2+h}-1\right) \left(e^{f r^2+h}+1\right)^2}\\
&&\hspace{-2.7cm}\phi_{19}=
\delta _1 \left(2 b^2 f r \text{csch}^2\left(f r^2+h\right)-4 b^2 f r^3 \coth \left(f r^2+h\right) \text{csch}^2\left(f r^2+h\right)\right) \left(8 b B_1 f r^2 \log (e) e^{f r^2+h}+\phi _2\right)\\
&&\hspace{-2.7cm} \phi_{20}=
A-\frac{b^4 r^2 \left(\delta _2 r^2-2 \delta _1\right) \text{csch}^4\left(f r^2+h\right)+\delta _2+\phi _1}{2 \phi _{15}^2}+\frac{\text{$\delta $1} \left(8 b B_1 f r^2 \log (e) e^{f r^2+h}+\phi _2\right)}{r^2 \phi _3 \left(e^{f r^2+h}-1\right) \left(e^{f r^2+h}+1\right)}\\
&&\hspace{-2.7cm} \phi_{21} = A+\frac{\delta _1 \left(8 b B_1 f r^2 \log (e) e^{f r^2+h}+\phi _2\right)}{r^2 \phi _3 \left(e^{f r^2+h}-1\right) \left(e^{f r^2+h}+1\right)}\\
&&\hspace{-2.7cm} \phi_{22}= \frac{2 f r \log (e) e^{f r^2+h}}{e^{f r^2+h}+1}-\frac{2 f r \log (e) e^{f r^2+h} \left(e^{f r^2+h}-1\right)}{\left(e^{f r^2+h}+1\right)^2}
\end{eqnarray*}

\begin{eqnarray*}
&&\hspace{-2.3cm} \phi_{23}= \frac{b B_1 \phi _{22} \left(e^{f r^2+h}+1\right) \left(e^{2 \left(f r^2+h\right)}-1\right)}{e^{f r^2+h}-1}\\
&&\hspace{-2.3cm} \phi_{24} = \frac{2 \delta _1 \left(8 b B_1 f r^2 \log (e) e^{f r^2+h}+\phi _2\right)}{r^3 \phi _3 \left(e^{b r^2+h}+1\right) \left(e^{f r^2+h}-1\right)}\\
&&\hspace{-2.3cm} \phi_{25}=\frac{4 f r \log (e) e^{2 \left(f r^2+h\right)} \left(2 A_1 f+b B_1 \log \left(\frac{e^{b r^2+c}-1}{e^{b r^2+c}+1}\right)\right)}{r^2 \phi _3 \left(e^{f r^2+h}-1\right) \left(e^{f r^2+h}+1\right)}\\
&&\hspace{-2.3cm} \phi_{26}=b^4 r^2 \left(\delta _2 r^2-2 \delta _1\right) \text{csch}^4\left(f r^2+h\right)+2 b^2 \text{csch}^2\left(f r^2+h\right) \left(-3 \delta _1+4 \delta _1 f r^2 \coth \left(f r^2+h\right)+\delta _2 r^2\right)+\delta _2\\
&&\hspace{-2.3cm} \phi_{27}=2 b^4 r \left(\delta _2 r^2-2 \delta _1\right) \text{csch}^4\left(f r^2+h\right)+2 b^4 \delta _2 r^3 \text{csch}^4\left(f r^2+h\right)\nonumber\\&&\hspace{-1.5cm}-8 b^4 f r^3 \left(\delta _2 r^2-2 \delta _1\right) \coth \left(f r^2+h\right) \text{csch}^4\left(f r^2+h\right)\nonumber\\&&\hspace{-1.5cm}+2 b^2 \text{csch}^2\left(f r^2+h\right) \left(-8 \delta _1 f^2 r^3 \text{csch}^2\left(f r^2+h\right)+8 \delta _1 f r \coth \left(f r^2+h\right)+2 \delta _2 r\right)\nonumber\\&&\hspace{-1.5cm}-8 b^2 f r \coth \left(f r^2+h\right) \text{csch}^2\left(f r^2+h\right) \left(-3 \delta _1+4 \delta _1 f r^2 \coth \left(f r^2+h\right)+\delta _2 r^2\right)\\
&&\hspace{-2.3cm} \phi_{28}=2 b^2 r \text{csch}^2\left(f r^2+h\right)-4 b^2 f r^3 \coth \left(f r^2+h\right) \text{csch}^2\left(f r^2+h\right)\\
&&\hspace{-2.3cm} \phi_{29}=
-\frac{\phi _{19}}{r^2 \phi _7 \left(e^{f r^2+h}-1\right) \left(e^{f r^2+h}+1\right)}\\
&&\hspace{-2.3cm} \phi_{30}=\frac{2 \delta _1}{r^3}-\frac{\phi _{16}}{r^2 \phi _5 \phi _{15}}-2 \phi _{18}-\phi _{24}-\phi _{29}\\
&&\hspace{-2.3cm} \Gamma =\frac{\phi _{20} \phi _{28} \phi _{26} \left(\delta _1 \left(16 b B_1 f^2 r^3 \log ^2(e) e^{f r^2+h}+16 b B_1 f r \log (e) e^{f r^2+h}+\phi _{23}+\phi _{25}\right)+\phi _{30}\right)}{\phi _{21} \phi _{15}^3}-\frac{\phi _{27}}{2 \phi _{15}^2}\\
&&\hspace{-2.3cm}\rho +p_{r}= A-\frac{b^4 r^2 \left(\delta _2 r^2-2 \delta _1\right) \text{csch}^4\left(f r^2+h\right)+\delta _2+\phi _1}{2 \phi _{15}^2}+\frac{\delta _1 \left(8 b B_1 f r^2 \log (e) e^{f r^2+h}+\phi _2\right)}{r^2 \phi _3 \left(e^{f r^2+h}-1\right) \left(e^{f r^2+h}+1\right)}\\
&&\hspace{-2.3cm}W_{1}=b^4 \delta _2 r^4 \phi _5 \text{csch}^4\left(f r^2+h\right)-16 b B_1 \delta _1 f^2 r^2 \log ^2(e) \left(e^{2 \left(f r^2+h\right)}+1\right) e^{f r^2+h}\nonumber\\&&\hspace{-1.8cm}+16 b B_1 \delta _1 f \log (e) \left(e^{2 \left(f r^2+h\right)}-1\right) e^{f r^2+h}+\delta _2 \phi _5\\
&&\hspace{-2.3cm}W_{2} = -8 b B_1 \delta _1 f^2 r^4 \log ^2(e) \left(e^{2 \left(f r^2+h\right)}+1\right) e^{f r^2+h}\nonumber\\&&\hspace{-1.8cm}+2 \delta _1 f r^2 \left(e^{2 \left(f r^2+h\right)}-1\right) \coth \left(f r^2+h\right) \left(4 b B_1 f r^2 \log (e) e^{f r^2+h}+\phi _2\right)\nonumber\\&&\hspace{-1.8cm}+4 b B_1 \delta _1 f r^2 \log (e) \left(e^{2 \left(f r^2+h\right)}-1\right) e^{f r^2+h}-\phi _5 \left(\delta _1-\delta _2 r^2\right)\\
&&\hspace{-2.3cm} \rho +p_{t} = -\frac{b^4 r^2 \left(\delta_{2} r^2-2 \delta_{1}\right) \text{csch}^4\left(b r^2+c\right)+\delta_{2}+\phi _1}{2 \phi _{15}^2}+\frac{2 b^2 W_2 \text{csch}^2\left(f r^2+h\right)}{2 \phi _7 \left(e^{b r^2+c}-1\right)^2 \left(e^{b r^2+c}+1\right)^2}+W_1\\
&&\hspace{-2.3cm}\rho+p_{r}+2p_{t} = -\frac{b^4 r^2 \left(\delta_{2} r^2-2 \delta_{1}\right) \text{csch}^4\left(f r^2+h\right)+\delta_{2}+\phi _1}{2 \phi _{15}^2}+\frac{2 b^2 W_2 \text{csch}^2\left(f r^2+h\right)}{\phi _7 \left(e^{f r^2+h}-1\right)^2 \left(e^{f r^2+h}+1\right)^2}\nonumber\\&&\hspace{-0.3cm}+\frac{\delta_{1} \left(8 b f \text{B1} r^2 \log (e) e^{f r^2+h}+\phi _2\right)}{r^2 \phi _3 \left(e^{f r^2+h}-1\right) \left(e^{f r^2+h}+1\right)}+A+W_1\\
&&\hspace{-2.3cm} \rho -p_{r} = -\frac{b^4 r^2 \left(\delta _2 r^2-2 \delta _1\right) \text{csch}^4\left(f r^2+h\right)+\delta _2+\phi _1}{2 \phi _{15}^2}+A-\frac{\delta _1 \left(8 b B_1 f r^2 \log (e) e^{f r^2+h}+\phi _2\right)}{r^2 \phi _3 \left(e^{f r^2+h}-1\right) \left(e^{f r^2+h}+1\right)}\\
&&\hspace{-2.3cm} \rho -p_{t} = -\frac{b^4 r^2 \left(\delta _2 r^2-2 \delta _1\right) \text{csch}^4\left(f r^2+h\right)+\delta _2+\phi _1}{2 \phi _{15}^2}+A+\frac{2 b^2 W_2 \text{csch}^2\left(f r^2+h\right)}{\phi _7 \left(e^{f r^2+h}-1\right)^2 \left(e^{fr^2+h}+1\right)^2}\nonumber\\&&\hspace{-1.4cm}+\frac{\delta _1 \left(8 b B_1 f r^2 \log (e) e^{f r^2+h}+\phi _2\right)}{r^2 \phi _3 \left(e^{f r^2+h}-1\right) \left(e^{f r^2+h}+1\right)}+W_1\\
&&\hspace{-2.3cm}W_{3}=b^4 \delta _2 r^4 \phi _5 \text{csch}^4\left(f r^2+h\right)-16 b B_1 \delta _1 f^2 r^2 \log ^2(e) \left(e^{2 \left(f r^2+h\right)}+1\right) e^{f r^2+h}\nonumber\\&&\hspace{-1.8cm}-16 b B_1 \delta _1 f \log (e) \left(e^{2 \left(f r^2+h\right)}-1\right) e^{f r^2+h}+\delta _2 \phi _5\\
&&\hspace{-2.3cm}\rho-p_{r}-2p_{t}=A-\frac{b^4 r^2 \left(\delta _2 r^2-2 \delta _1\right) \text{csch}^4\left(f r^2+h\right)+\delta _2+\phi _1}{2 \phi _{15}^2}+b^4 \delta _2 r^4 \phi _5 \text{csch}^4\left(f r^2+h\right)\nonumber\\&&\hspace{-0.3cm}+\frac{2 b^2 W_2 \text{csch}^2\left(f r^2+h\right)}{\phi _7 \left(e^{f r^2+h}-1\right)^2 \left(e^{f r^2+h}+1\right)^2}-\frac{\text{$\delta $1} \left(8 b B_1 f r^2 \log (e) e^{f r^2+h}+\phi _2\right)}{r^2 \phi _3 \left(e^{f r^2+h}-1\right) \left(e^{f r^2+h}+1\right)}-W_3
\end{eqnarray*}

\begin{eqnarray*}
&&\hspace{-0.5cm}Z_{1} = A-\frac{b^4 r^2 \left(\delta _2 r^2-2 \delta _1\right) \text{csch}^4\left(f r^2+h\right)+\delta _2+\phi _1}{2 \phi _{15}^2}+\frac{\phi _{24}}{2}\\
&&\hspace{-0.5cm}Z_{2} = \frac{2 \delta _1 f \log (e) e^{f r^2+h} \left(8 b B_1 f r^2 \log (e) e^{f r^2+h}+\phi _2\right)}{r \phi _3 \left(e^{f r^2+h}-1\right)^2 \left(e^{f r^2+h}+1\right)}\\
&&\hspace{-0.5cm}Z_{3} = 16 b B_1 f^2 r^3 \log ^2(e) e^{f r^2+h}+16 b B_1 f r \log (e) e^{f r^2+h}\\
&&\hspace{-0.5cm}Z_{4}=\left(A_1+\frac{b B_1 \log \left(\frac{e^{f r^2+h}-1}{e^{f r^2+h}+1}\right)}{2 f}\right)\\
&&\hspace{-0.5cm}Z_{5} =\frac{b B_1 \phi _{22} \left(e^{f r^2+h}+1\right) \left(e^{2 \left(f r^2+h\right)}-1\right)}{e^{f r^2+h}-1}\\
&&\hspace{-0.5cm}Z_{6} = 2 b^4 \delta _2 r^3 \text{csch}^4\left(f r^2+h\right)-8 b^2 f r \coth \left(f r^2+h\right) \text{csch}^2\left(f r^2+h\right) \left(-3 \delta _1+4 \delta _1 f r^2 \coth \left(f r^2+h\right)+\delta _2 r^2\right)\\
&&\hspace{-0.5cm}Z_{7} = 2 b^4 r \left(\delta _2 r^2-2 \delta _1\right) \text{csch}^4\left(f r^2+h\right)-8 b^4 f r^3 \left(\delta _2 r^2-2 \delta _1\right) \coth \left(f r^2+h\right) \text{csch}^4\left(f r^2+h\right)\nonumber\\&&\hspace{0.0cm}+2 b^2 \text{csch}^2\left(f r^2+h\right) \left(-8 \delta _1 f^2 r^3 \text{csch}^2\left(f r^2+h\right)+8 \delta _1 f r \coth \left(f r^2+h\right)+2 \delta _2 r\right)\\
&&\hspace{-0.5cm}Z_{8} = -8 b^2 f \coth \left(f r^2+h\right) \text{csch}^2\left(f r^2+h\right) \left(-3 \delta _1+4 \delta _1 f r^2 \coth \left(f r^2+h\right)+\delta _2 r^2\right)\\
&&\hspace{-0.5cm}Z_{9} = 32 b^2 f^2 r^2 \coth ^2\left(f r^2+h\right) \text{csch}^2\left(f r^2+h\right) \left(-3 \delta _1+4 \delta _1 f r^2 \coth \left(f r^2+h\right)+\delta _2 r^2\right)\\
&&\hspace{-0.5cm}Z_{10} = 10 b^4 \delta _1 r^2 \text{csch}^4\left(f r^2+h\right)+2 b^4 \left(\delta _2 r^2-2 \delta _1\right) \text{csch}^4\left(f r^2+h\right)-32 b^4 \delta _2 f r^4 \coth \left(f r^2+h\right) \text{csch}^4\left(f r^2+h\right)\\
&&\hspace{-0.5cm}Z_{11} = 16 b^4 f^2 r^4 \left(\delta _2 r^2-2 \delta _1\right) \text{csch}^6\left(f r^2+h\right)+64 b^4 f^2 r^4 \left(\delta _2 r^2-2 \delta _1\right) \coth ^2\left(f r^2+h\right) \text{csch}^4\left(f r^2+h\right)\nonumber\\&&\hspace{0.1cm}+40 b^4 f r^2 \left(\delta _2 r^2-2 \delta _1\right) \coth \left(f r^2+h\right) \text{csch}^4\left(f r^2+h\right)+16 b^2 f^2 r^2 \text{csch}^4\left(f r^2+h\right) \left(4 b \delta _1 r^2 \coth \left(f r^2+h\right)-3 \delta _1+\delta _2 r^2\right)\\
&&\hspace{-0.5cm}Z_{12} = 16 b^2 f r \coth \left(f r^2+h\right) \text{csch}^2\left(f r^2+h\right) \left(-8 \delta _1 f^2 r^3 \text{csch}^2\left(b r^2+c\right)+8 \delta _1 f r \coth \left(f r^2+h\right)+2 \delta _1 r\right)\\
&&\hspace{-0.5cm}Z_{13} = 2 b^2 \text{csch}^2\left(f r^2+h\right) \left(2 \delta _2-40 \delta _2 f^2 r^2 \text{csch}^2\left(f r^2+h\right)+8 \delta _1 f \coth \left(f r^2+f\right)\right)\\
&&\hspace{-0.5cm}Z_{14}=b^4 r^2 \left(\delta _2 r^2-2 \delta _1\right) \text{csch}^4\left(f r^2+h\right)+\delta _2+\phi _1\\
&&\hspace{-0.5cm}Z_{15}=8 b^2 f^2 r^4 \text{csch}^4\left(f r^2+h\right)+16 b^2 f^2 r^4 \coth ^2\left(f r^2+h\right) \text{csch}^2\left(f r^2+h\right)+2 b^2 \text{csch}^2\left(f r^2+h\right)-20 b^2 f r^2 \coth \left(f r^2+h\right) \nonumber\\&& \hspace{0.1cm} \times \text{csch}^2\left(f r^2+h\right)
\\
&&\hspace{-0.5cm}F_{g} = -\frac{b B_1 Z_1 \phi _{22} \left(e^{f r^2+h}+1\right)}{2 f \left(e^{f r^2+h}-1\right) Z_{4}}\\
&&\hspace{-0.5cm}F_{h} =\frac{a B_1 \delta _1 \phi _{22} \left(8 b B_1 f r^2 \log (e) e^{f r^2+h}+\phi _2\right)}{r^2 \phi _2 \phi _5}\nonumber\\&&\hspace{0.0 cm}+\frac{\delta _1 \left(2 b^2 r \text{csch}^2\left(f r^2+h\right)-4 b^2 b r^3 \coth \left(f r^2+h\right) \text{csch}^2\left(f r^2+h\right)\right) \left(8 b B_1 f r^2 \log (e) e^{f r^2+h}+\phi _2\right)}{r^2 \phi _7 \left(e^{f r^2+h}-1\right) \left(e^{f r^2+h}+1\right)}\nonumber\\&&\hspace{0.0cm}-\frac{\delta _1 \left(4 f r Z_4 \log (e) e^{2 \left(f r^2+h\right)}+Z_3+Z_5\right)}{r^2 \phi _3 \left(e^{f r^2+h}-1\right) \left(e^{f r^2+h}+1\right)}-\frac{2 \delta _1}{r^3}+Z_2+\phi _{18}+\phi _{24}\\
&&\hspace{-0.5cm}F_{a} =\frac{2}{r}\left(\frac{2 a^2 W_2 \text{csch}^2\left(b r^2+c\right)}{2 \phi _7 \left(e^{b r^2+c}-1\right)^2 \left(e^{b r^2+c}+1\right)^2}+A-\frac{r \phi _{24}}{2}+W_1\right)\\
&&\hspace{-0.5cm}\rho{\prime \prime}=\frac{- Z_{15} \left(a^4 r^2 \left(\text{$\delta $2} r^2-2 \text{$\delta $1}\right) \text{csch}^4\left(b r^2+c\right)+\text{$\delta $2}+\phi _1\right)}{\phi _{15}^7}\nonumber\\&&\hspace{0.0cm}-\frac{3 Z_{14} \left(2 a^2 r \text{csch}^2\left(b r^2+c\right)-4 a^2 b r^3 \coth \left(b r^2+c\right) \text{csch}^2\left(b r^2+c\right)\right)^2}{\phi _{15}^4}\nonumber\\&&\hspace{0.0cm}-\frac{2 \left(Z_6+Z_7\right) \left(2 a^2 r \text{csch}^2\left(b r^2+c\right)-4 a^2 b r^3 \coth \left(b r^2+c\right) \text{csch}^2\left(b r^2+c\right)\right)}{\phi _{15}^7}\nonumber\\&&\hspace{0.0cm}+\frac{Z_8+Z_9+Z_{10}-Z_{11}-Z_{12}+Z_{13}+Z_{14}}{2 \phi _{15}^6}\\
&&\hspace{-0.5cm}T_{1} = b^3 \left(e^{2 \left(f r^2+h\right)}-1\right)^2 \text{csch}^4\left(f r^2+h\right) \left(2 A_1 f+b B_1 \log \left(\frac{e^{f r^2+h}-1}{e^{f r^2+h}+1}\right)\right)^2\\
\end{eqnarray*}

\begin{eqnarray*}
&&\hspace{-0.7cm}T_{2} = 2 A_1 f \left(e^{2 \left(f r^2+h\right)}+1\right)+2 b B_1 e^{f r^2+h}+b B_1 \left(e^{2 \left(f r^2+h\right)}+1\right) \log \left(\frac{e^{f r^2+h}-1}{e^{f r^2+h}+1}\right)\\
&&\hspace{-0.7cm}T_{3} = 
2 b f \phi _2 \coth \left(f r^2+h\right) \text{csch}^2\left(f r^2+h\right) \left(8 b B_1 f r^2 \log (e) e^{f r^2+h}+\phi _2\right)\\
&&\hspace{-0.7cm}T_{4} = f r^2 \log (e) \left(2 A_1 f \left(e^{2 \left(f r^2+h\right)}+1\right)+2 b B_1 e^{f r^2+h}+b B_1 \left(e^{2 \left(f r^2+h\right)}+1\right) \log \left(\frac{e^{f r^2+h}-1}{e^{f r^2+h}+1}\right)\right)+\phi _2\\
&&\hspace{-0.7cm}T_{5} =\text{csch}^2\left(f r^2+h\right) \left(b^2 r^2 \left(b^2-4 f^2 r^2\right) \text{csch}^2\left(f r^2+h\right)+5 b^2-4 f^2 r^2\right)\nonumber\\&&\hspace{-0.2cm}+8 f^2 r^2 \coth ^2\left(f r^2+h\right) \left(b^2 r^2 \text{csch}^2\left(f r^2+h\right)-1\right)+2 f \coth \left(f r^2+h\right) \left(5-3 b^2 r^2 \text{csch}^2\left(f r^2+h\right)\right)\\
&&\hspace{-0.7cm}D_{1} = b^4 \delta _2 r^4 \phi _5 \text{csch}^4\left(f r^2+h\right)-16 b^3 B_1 \delta _1 f^2 r^4 \log ^2(e) \left(e^{2 \left(f r^2+h\right)}+1\right) e^{f r^2+h} \text{csch}^2\left(f r^2+h\right)\nonumber\\&&\hspace{-0.2cm}+4 b^2 \delta _1 f r^2 \left(e^{2 \left(f r^2+h\right)}-1\right) \coth \left(f r^2+h\right) \text{csch}^2\left(f r^2+h\right) \left(4 b B_1 f r^2 \log (e) e^{f r^2+h}+\phi _2\right)\nonumber\\&&\hspace{-0.2cm}+8 b b^2 B_1 \delta _1 f r^2 \log (e) \left(e^{2 \left(f r^2+h\right)}-1\right) e^{f r^2+h} \text{csch}^2\left(f r^2+h\right)-2 b^2 \phi _5 \left(\delta _1-\delta _2 r^2\right) \text{csch}^2\left(f r^2+h\right)\nonumber\\&&\hspace{-0.2cm}-16 b B_1 \delta _1 f^2 r^2 \log ^2(e) \left(e^{2 \left(f r^2+h\right)}+1\right) e^{f r^2+h}+16 b B_1 \delta _1 f \log (e) \left(e^{2 \left(f r^2+h\right)}-1\right) e^{f r^2+h}+\delta _2 \phi _5\\
&&\hspace{-0.7cm}D_{2} = 4 b^5 B_1 \delta _2 f r^5 \log (e) \left(-e^{f r^2+h}\right) \left(e^{f r^2+h}+1\right) e^{f r^2+h} \text{csch}^4\left(f r^2+h\right)+8 b^4 \delta _2 f r^5 \phi _2 \log (e) e^{2 \left(f r^2+h\right)} \text{csch}^4\left(f r^2+h\right)\nonumber\\&&\hspace{-0.2cm}-8 b^4 \delta _2 f r^5 \phi _5 \coth \left(f r^2+h\right) \text{csch}^4\left(f r^2+h\right)+4 b^4 \delta _2 r^3 \phi _5 \text{csch}^4\left(f r^2+h\right)\nonumber\\&&\hspace{-0.2cm}-64 b^3 B_1 \delta _1 f^3 r r^4 \log ^2(e) \left(e^{2 \left(f r^2+h\right)}+1\right) e^{f r^2+h} \coth \left(f r^2+h\right) \text{csch}^2\left(f r^2+h\right)\nonumber\\&&\hspace{-0.2cm}-32 b^3 B_1 \delta _1 f^2 r r^2 \log (e) \left(e^{2 \left(f r^2+h\right)}-1\right) e^{f r^2+h} \coth \left(f r^2+h\right) \text{csch}^2\left(f r^2+h\right)\nonumber\\&&\hspace{-0.2cm}+16 b^2 \delta _1 f^2 r^3 \left(e^{2 \left(f r^2+h\right)}-1\right) \coth ^2\left(f r^2+h\right) \text{csch}^3\left(f r^2+h\right) \left(4 b B_1 f r^2 \log (e) e^{f r^2+h}+\phi _2\right)\nonumber\\&&\hspace{-0.2cm}-8 b^2 f r \phi _5 \left(\delta _1-\delta _2 r^2\right) \coth \left(f r^2+h\right) \text{csch}^2\left(f r^2+h\right)-64 b B_1 \delta _1 f^3 r^3 \log ^3(e) e^{3 \left(b r^2+c\right)}\nonumber\\&&\hspace{-0.2cm}-32 b B_1 \delta _1 f^3 r^3 \log ^3(e) \left(e^{2 \left(f r^2+h\right)}+1\right) e^{f r^2+h}-32 b B_1 \delta _1 f^2 r \log ^2(e) \left(e^{2 \left(f r^2+h\right)}+1\right) e^{f r^2+h}\nonumber\\&&\hspace{-0.2cm}+32 b B_1 \delta _1 f^2 r \log ^2(e) \left(e^{2 \left(f r^2+h\right)}-1\right) e^{f r^2+h}+64 b B_1 \delta _1 f^2 r \log ^2(e) e^{3 \left(f r^2+h\right)}\nonumber\\&&\hspace{-0.2cm}+4 b B_1 \delta _2 f r \log (e) \left(e^{f r^2+h}-1\right) \left(e^{f r^2+h}+1\right) e^{f r^2+h}+8 b \delta _2 r \phi _2 \log (e) e^{2 \left(f r^2+h\right)}\\
&&\hspace{-0.7cm}D_{3}=8 \delta _1 f^2 r^3 \log (e) e^{f r^2+h} \left(e^{2 \left(f r^2+h\right)}-1\right) \coth \left(f r^2+h\right)\\
&&\hspace{-0.7cm}D_{4}=D_{3} \left(2 A_1 f e^{f r^2+h}+b B_1 e^{f r^2+h} \log \left(\frac{e^{f r^2+h}-1}{e^{f r^2+h}+1}\right)+2 b B_1 f r^2 \log (e)+3 B_1 b\right)\nonumber\\&&\hspace{-0.2cm}-16 b B_1 \delta _1 f^3 r^5 \log ^3(e) \left(e^{2 \left(f r^2+h\right)}+1\right) e^{f r^2+h}-32 b B_1 \delta _1 f^3 r^5 \log ^3(e) e^{3 \left(f r^2+h\right)}\nonumber\\&&\hspace{-0.2cm}+8 \delta _1 f^2 r^3 \log (e) e^{2 \left(f r^2+h\right)} \coth \left(f r^2+h\right) \left(4 b B_1 f r^2 \log (e) e^{f r^2+h}+\phi _2\right)\nonumber\\&&\hspace{-0.2cm}-4 \delta _1 f^2 r^3 \left(e^{2 \left(f r^2+h\right)}-1\right) \text{csch}^2\left(f r^2+h\right) \left(4 b B_1 f r^2 \log (e) e^{f r^2+h}+\phi _2\right)\nonumber\\&&\hspace{-0.2cm}-32 b B_1 \delta _1 f^2 r^3 \log ^2(e) \left(e^{2 \left(f r^2+h\right)}+1\right) e^{f r^2+h}+8 b B_1 \delta _1 f^2 r^3 \log ^2(e) \left(e^{2 \left(f r^2+h\right)}-1\right) e^{f r^2+h}\nonumber\\&&\hspace{-0.2cm}+16 b B_1 \delta _1 f^2 r^3 \log ^2(e) e^{3 \left(f r^2+h\right)}+4 \delta _1 f r \left(e^{2 \left(f r^2+h\right)}-1\right) \coth \left(f r^2+h\right) \left(4 b B_1 f r^2 \log (e) e^{f r^2+h}+\phi _2\right)\nonumber\\&&\hspace{-0.2cm}+8 b B_1 \delta _1 f r \log (e) \left(e^{2 \left(f r^2+h\right)}-1\right) e^{f r^2+h}+4 b B_1 f r \log (e) \left(\delta _2 r^2-\delta _1\right) \left(e^{f r^2+h}-1\right) \left(e^{f r^2+h}+1\right) e^{f r^2+h}\nonumber\\&&\hspace{-0.2cm}+8 f r \phi _2 \log (e) \left(\text{$\delta $2} r^2-\text{$\delta $1}\right) e^{2 \left(b r^2+c\right)}+2 \delta _2 r \phi _5\\
&&\hspace{-0.7cm}D_{5} = -4 b^4 r^2-2 \left(b^2 \left(8 f^2 r^4+5\right)-4 f^2 r^2\right) \cosh \left(2 \left(f r^2+h\right)\right)+12 b^2 f r^2 \sinh \left(2 \left(f r^2+h\right)\right)+10 b^2\nonumber\\&&\hspace{-0.2cm}-5 f \sinh \left(4 \left(b r^2+c\right)\right)+4 f^2 r^2 \cosh \left(4 \left(f r^2+h\right)\right)-12 f^2 r^2+10 f \sinh \left(2 \left(f r^2+h\right)\right)\\
&&\hspace{-0.7cm}D_{6}=4 b^2 r \left(e^{f r^2+h}-1\right) \left(e^{f r^2+h}+1\right) \left(1-2 f r^2 \coth \left(f r^2+h\right)\right) \text{csch}^2\left(f r^2+h\right)
\\
&&\hspace{-0.7cm}V^{2}_{r} = -\frac{\phi _{15} \sinh ^2\left(f r^2+h\right) \left(-8 b^2 f B_{1} T_4 \log (e) e^{f r^2+h} \text{csch}^2\left(f r^2+h\right)-8 b^2 B_{1} T_2 \log ^2(e) e^{f r^2+h}+T_1+T_3\right)}{a T_5 \left(e^{f r^2+h}-1\right)^2 \left(e^{f r^2+h}+1\right)^2 \left(f B_{1} \log \left(\frac{e^{f r^2+h}-1}{e^{f r^2+h}+1}\right)+2 A_{1} b\right)^2}\\
\end{eqnarray*}
\begin{eqnarray*}
&&\hspace{-1.6cm}V^{2}_{t} =\frac{\phi _3 \left(e^{\text{bf} r^2+h}-1\right) \left(e^{\text{bf} r^2+h}+1\right) \sinh ^6\left(\text{bf} r^2+h\right) \left(2 b^2 D_3 \text{csch}^2\left(\text{bf} r^2+h\right)+D_2\right)}{b^2 \delta _1 D_5 r \left(e^{2 \left(\text{bf} r^2+h\right)}-1\right)^3 \left(2 A_1 f+b B_1 \log \left(\frac{e^{\text{bf} r^2+h}-1}{e^{\text{bf} r^2+h}+1}\right)\right){}^2}\nonumber\\&&\hspace{-1.1cm}-4 D_1 f r \phi _{15} \log (e) \left(e^{\text{bf} r^2+h}+1\right) e^{\text{bf} r^2+h} \sinh ^6\left(\text{bf} r^2+h\right) \left(2 A_1 f+b B_1 \log \left(\frac{e^{\text{bf} r^2+h}-1}{e^{\text{bf} r^2+h}+1}\right)\right)\nonumber\\&&\hspace{-1.1cm}-D_6 D_1 \sinh ^6\left(\text{bf} r^2+h\right) \left(2 A_1 f+b B_1 \log \left(\frac{e^{\text{bf} r^2+h}-1}{e^{\text{bf} r^2+h}+1}\right)\right)-4 b B_1 D_1 f r \phi _{15} \log (e) e^{\text{bf} r^2+h} \sinh ^6\left(\text{bf} r^2+h\right)\nonumber\\&&\hspace{-1.1cm}-4 D_1 f r \phi _3 \log (e) \left(e^{\text{bf} r^2+h}-1\right) e^{\text{bf} r^2+h} \sinh ^6\left(\text{bf} r^2+h\right)
\\
\end{eqnarray*}
\end{widetext}

\end{document}